\PassOptionsToPackage{dvipsnames}{xcolor} % prevent an option clash
\documentclass{sig-alternate}
\usepackage{mathtools}
\usepackage{graphicx}
\usepackage{subfigure}

\usepackage{amsthm}
\usepackage{color}
\usepackage{multirow}
\usepackage{xspace}
\usepackage{appendix}
\usepackage{balance}	
\usepackage{algorithm}	
\usepackage{algorithmicx}
\usepackage{algpseudocode}
\usepackage{ellipsis}
\usepackage{url}
\usepackage{subfigure}
\usepackage{colortbl}
\usepackage[table]{xcolor}
\usepackage[dvipsnames]{xcolor}
\usepackage{color,soul} 
\usepackage{enumitem}
\usepackage{tabularx}
\usepackage{amsmath}
\usepackage{tikz}
\usetikzlibrary{positioning,shapes.geometric}
\algrenewcommand\algorithmicindent{0.5em}

\definecolor{Mygrey}{gray}{0.75}
\definecolor{MyGray}{rgb}{0.85,0.85,0.85}

\newcommand{\SetRowColor}[1]{\noalign{\gdef\RowColorName{#1}}\rowcolor{\RowColorName}}

\definecolor{MyGray}{rgb}{0.85,0.85,0.85}

\newcommand{\eg}{{\em e.g.}}
\newcommand{\ie}{{\em i.e.}}

\newcommand{\eat}[1]{}
\newcommand{\cut}[1]{}

\newcommand*{\rom}[1]{\expandafter\@slowromancap\romannumeral #1@}

\theoremstyle{definition}
\newtheorem{definition}{Definition}[section]
\newtheorem{example}{Example}[section]

\newcommand\xqed[1]{%
  \leavevmode\unskip\penalty9999 \hbox{}\nobreak\hfill
  \quad\hbox{#1}}
\newcommand\marker{\xqed{$\square$}}

\makeatletter
\newcommand{\algcolor}[2]{%
  \hskip-\ALG@thistlm\colorbox{#1}{\parbox{\dimexpr\linewidth-2\fboxsep}{\hskip\ALG@thistlm\relax #2}}%
}

\newcounter{parentalgorithm}

\makeatother

\algnewcommand\algorithmicswitch{\textbf{switch}}
\algnewcommand\algorithmiccase{\textbf{case}}
\algnewcommand\algorithmicdefault{\textbf{default}}
\algnewcommand\algorithmicassert{\texttt{assert}}
\algnewcommand\Assert[1]{\State \algorithmicassert(#1)}%
\algdef{SE}[SWITCH]{Switch}{EndSwitch}[1]{\algorithmicswitch\ #1\ \algorithmicdo}{\algorithmicend\ \algorithmicswitch}%
\algdef{SE}[CASE]{Case}{EndCase}[1]{\algorithmiccase\ #1}{\algorithmicend\ \algorithmiccase}%
\algdef{SE}[CASE]{Default}{EndDefault}[1]{\algorithmicdefault\ #1}{\algorithmicend\ \algorithmicdefault}%
\algtext*{EndSwitch}%
\algtext*{EndCase}%
\algtext*{EndDefault}%

\begin{document}
%\conferenceinfo{SIGMOD'11,} {June 12--16, 2011, Athens, Greece.}
%\CopyrightYear{2011}
%\crdata{978-1-4503-0661-4/11/06}
%\clubpenalty=10000
%\widowpenalty = 10000
%\title{\name: A Scientific Array Database Travels in a Time Machine}

\title{Efficient Iterative Processing \\ in the SciDB Parallel Array Engine}

\numberofauthors{1} 
\author{
  % 1st. author
 \alignauthor
 Emad Soroush$^1$, Magdalena Balazinska$^1$, Simon Krughoff$^2$, and Andrew Connolly$^2$\\
          \affaddr{$^1$Dept. of Computer Science \& Engineering ~~~~~ $^2$ Astronomy Department}\\
          \affaddr{University of Washington, Seattle, USA}\\
          \email{\{soroush,magda\}@cs.washington.edu}
          \email{\{krughoff, ajc\}@astro.washington.edu}
%   % 2nd. author
% \alignauthor 
% Magdalena Balazinska\\
 %         \affaddr{Dept. of Computer Science \& Engineering}\\
  %        \affaddr{University of Washington}\\
   %       \affaddr{Seattle, USA}\\
    %      \email{magda@cs.washington.edu}
}
      
\maketitle

\floatstyle{ruled}
\newfloat{Listing}{thp}{lop}
\floatname{Listing}{Listing}

\begin{sloppypar}

%\input{abstract}

%\cat egory{H.2.4}{Information Systems}{Database Management $-$$-$$-$}{Systems}
%\category{H.2.8}{Information Systems}{Database Management$-$$-$$-$}{Database applications}

%\vspace{-0.3cm}
%\terms{Algorithms, Design, Performance}

%\vspace{-0.3cm}
%\keywords{Overlap Execution Strategy, Parallel Databases, Query Processing, Scientific Databases}
\section*{Abstract}

Many scientific data-intensive applications perform iterative
computations on array data. There exist multiple engines specialized for
array processing. These engines efficiently support various types of
operations, but none includes native support for iterative
processing. In this paper, we develop a model for iterative array
computations and a series of optimizations. We evaluate the benefits of
an optimized, native support for iterative array processing on the
SciDB engine and real workloads from the astronomy domain.

\section{Introduction}
\label{sec:intro}

Science is increasingly becoming data-driven~\cite{hey:09}. From small
research labs to large communities~\cite{sdss,lsst},
scientists have access to more data than ever before. As a result,
scientists can increasingly benefit from using database
management systems to organize and query their
data~\cite{loebman:09,skyserver}.

%~\cite{Zoumpatianos:2014}.

Scientific data often takes the form of multidimensional arrays (\eg,
2D images or 3D environment simulations). One approach to managing
this type of array data is to build array libraries on top of
relational engines, but many argue that simulating arrays on top of
relations can be highly inefficient~\cite{curde:10}. Scientists also
need to perform a variety of operations on their array data such as
feature extraction~\cite{kwon:10b}, smoothing~\cite{rogers:10}, and
cross-matching~\cite{NietoSantisteban:06}, which are not built-in
operations in relational Database Management Systems (DBMSs).  Those
operations also impose different requirements than relational
operators on a DBMS~\cite{soroush:11a}.

As a result, many data management systems are being built to support
the array model natively~\cite{baumann:98,rogers:10,zhang:09}.
Additionally, to handle today's large-scale datasets, several engines,
including SciDB~\cite{rogers:10}, provide support for processing
arrays in parallel in a shared-nothing cluster.  Several benchmark
studies have shown that these specialized array engines outperform
both relational engines and MapReduce-type systems on a variety of
array workloads~\cite{curde:10,taft:14}.

%In array-based systems, structural
%information is associated with each cell through its dimension values.
%Additionally, array dimensions provide a natural index for the data,
%which improves query performance.

%is an open-source parallel database management system 
%where the core data model is a multi-dimensional array. SciDB supports arrays \textit{natively}. 
%That is, SciDB is designed and implemented from the ground-up based on array data model rather than being 
%layered on top of an existing DBMS~\cite{ballegooij:05,cohen:09,baumann:98,zhang:09}.

%To support these growing data management
%needs many advocate that one should move away from the relational
%model and adopt a multidimensional array data model~\cite{baumann:98,stonebraker:09}. 
%The main reason is that scientists typically work with multidimensional measurements 
%such as arrays and meshes that their data values are associated with coordinates in space and time. 
%Furthermore, simulating arrays on top of relations can be highly
%inefficient~\cite{stonebraker:09}. Scientists also need to perform
%array-specific operations such as feature extraction~\cite{kwon:10b},
%smoothing~\cite{rogers:10}, and cross-matching~\cite{NietoSantisteban:06}, which
%are not built-in operations in relational DBMSs. 

Many data analysis tasks today require iterative
processing~\cite{ewen:12}: machine learning, model fitting, pattern
discovery, flow simulations, cluster extraction, and more. As a
result, most modern Big Data management and analytics systems
(\eg,~\cite{low:12,zaharia:10}) support iterative processing as a
first-class citizen and offer a variety of optimizations for these
types of computations: caching~\cite{bu:10}, asynchronous
processing~\cite{low:12}, prioritized
processing~\cite{mcsherry:13,zhang:11}, etc.

The need for efficient iterative computation extends to analysis
executed on multi-dimensional scientific arrays. For example,
astronomers typically apply an iterative outlier-removal algorithm to
telescope images as one of the first data processing steps.  Once the
telescope images have been cleaned, the next processing step is to
extract sources (\ie, stars, galaxies, and other celestial structures)
from these images. The source extraction algorithm is most easily
written as an iterative query as well. As a third example, the
simple task of clustering data in a multi-dimensional array also
requires iterating until convergence to the final set of clusters. 
We further describe these three applications in Section~\ref{sec:examples}.

% incremental iterative processing 

While it is possible to implement iterative array computations by
repeatedly invoking array queries from a script, this approach is
highly inefficient (as we show in
Figure~\ref{fig:SigmaClip-incr}). Instead, a large-scale array
management systems such as SciDB should support iterative computations
as first-class citizens in the same way other modern data management
systems do for relational or graph data.

\noindent \textbf{Contributions:} In this paper, we introduce a new
model for expressing iterative queries over arrays. We develop a
middleware system called ArrayLoop that we implement on top of SciDB
to translate queries expressed in this model into queries that can be
executed in SciDB. Importantly, ArrayLoop includes three optimizations
that trigger rewrites to the iterative queries and ensure their
efficient evaluation. The first optimization also includes extensions
to the SciDB storage manager. More specifically, the contribution
of this paper are as follows:

\noindent \textbf{(1) New model for iterative array processing}
(Sections~\ref{sec:model} and~\ref{sec:exec}): Iterating over arrays is
different from iterating over relations.  In the case of arrays, the
iteration starts with an array and updates the cell values of that
array. It does not generate new tuples as in a relational query.
Additionally, these update operations typically operate on
neighborhoods of cells.  These two properties are the foundation of
our new model for iterative array processing. Our model enables the
declarative specification of iterative array computations, their
automated optimization, and their efficient execution.

%We further consider
%the following three optimizations that significantly speed-up
%iterative array computations:

\noindent \textbf{(2)Incremental iterative processing}
(Section~\ref{sec:inc}): In many iterative applications, the result of
the computation changes only partly from one iteration to the next. As
such, implementations that recompute the entire result every time are
known to be inefficient.  The optimization, called \textit{incremental
  iterative processing}~\cite{ewen:12}, involves processing only the
part of the data that changes across iterations. When this
optimization is applicable, it has been shown to significantly improve
performance in relational and graph
systems~\cite{ewen:12,mcsherry:13}.  This optimization also applies to
array iterations. While it is possible to manually write a set of
queries that process the data incrementally, doing so is tedious,
error-prone, and can miss optimization opportunities.  Our iterative
array model enables the automatic generation of such incremental
computations from the user's declarative specification of the overall
iterative query. Additionally, while the idea of incremental
iterations has previously been developed for relational systems, its
implementation in an array engine is very different: For an array
engine, the optimization can be pushed \textit{all the way to the
  storage manager} with significant performance benefits. We develop
and evaluate such storage-manager-based approach to incremental array
processing.

%To address this challenge, we develop an
%approach that enables users to specify their computation at a logical,
%high-level. The system automatically generates an incremental version
%of the computation. Additionally, we optimize the way in which the
%system handles all partial results and merges them with the full
%result at each iteration. We show that several of these optimizations
%can be pushed all the way down to the storage manager in an array
%system.
%The first optimization, which focuses on making
%iterations incremental, is based on analogous optimizations in
%relational and graph systems, but its 

%overlap processing 
\noindent \textbf{(3) Overlap iterative processing}
(Section~\ref{sec:overlap}): In iterative array applications,
including, for example, cluster finding and source detection,
operations in the body of the loop update the value of the array cells
by using the values of other neighboring array cells. These
neighborhoods are often bounded in size.  These applications can
effectively be processed in parallel if the system partitions an array
but also replicates a small amount of overlap cells. In the case of
iterative processing, the key challenge lies in keeping these overlap
cells up to date. This optimization is specific to queries over arrays
and does not apply to relational engines.  Our key contribution here
lies in new mechanisms for managing the efficient reshuffling of the
overlap data across iterations.

A subset of applications that leverage overlap data also have the
property that overlap cells can be updated only every few
iterations. Examples of such applications are those that try to find
structures in the array data. They can find structures locally, and
need to exchange information only periodically to stitch these local
structures into larger ones. We extend our overlap data shuffling
approach to leverage this property and further reduce the overhead of
synchronizing overlap data.  We call this optimization,
mini-iterations.

%The other two optimizations that we implement are specific to arrays:
%One optimization optimizes the processing of overlap cells when
%parallelizing an iterative array computation. The key contribution
%here lies in new mechanisms for managing the efficient reshuffling of
%the overlap data across iterations. 

% multi-resolution optimization
\noindent \textbf{(4)Multi-resolution iterative processing}
(Section~\ref{sec:multi-res}): Finally, in many applications, the raw
data lives in a continuous space (3D universe, 2D ocean, N-D space of
continuous variables) and arrays capture discretized approximations of
the real data. Different data resolutions are thus possible and
scientifically meaningful to analyze. In fact, it is common for
scientists to look at the data at different levels of detail. In many
applications, it is often efficient to first process the
low-resolution versions of the data and use the result to speed-up the
processing of finer-resolution versions of the data if requested by
the user. Our final optimization automates this approach. While
scientists are accustomed to working with arrays at different levels
or detail, our contribution is to show how this optimization can be
automatically applied to iterative queries in an array engine.

%The last optimization, leverages%
%multi-resolution array processing: it first processes low-resolution
%versions of the data to boost the performance on fine-resolution
%versions. 

%We call this technique
%\textit{multi-resolution iterative processing}. and show its
%benefit on the source detection application.

%\emad{list the contributions of this paper.}

\noindent \textbf{(5) Implementation and evaluation} We implement the
iterative model and all three optimizations as extensions to the
open-source SciDB engine and we demonstrate their effectiveness on
experiments with 1~TB of publically-available synthetic LSST
images~\cite{repository}. Experiments show that \textit{Incremental
  iterative processing} can boost performance by a factor of 4-6X
compared to a non-incremental iterative computation.
\textit{Iterative overlap processing} together with
\textit{mini-iteration processing} can improve performance by 31\%
compare to SciDB's current implementation of overlap
processing. Finally, the \textit{multi-resolution optimization} can cut
runtimes in half if an application can leverage this
technique. Interestingly, these three optimizations are complementary
and their benefits can be compounded.

To the best of our knowledge, this paper is
the first to design, implement, and evaluate an approach for iterative
processing in a parallel array data management system. Given that
array engines have been shown to outperform a variety of other systems
on array workloads~\cite{curde:10,taft:14} and that iterative
analytics are common on array data (as we discussed above), efficient
support for iterative query processing in array engines is a critical
component of the big data engine ecosystem. 

% The rest of this paper is organized as follows: We start with a more
% detailed description of three motivation applications in
% Section~\ref{sec:examples}. In Section~\ref{sec:model}, we formally
% define the iterative array model. In Section~\ref{sec:inc}, we present
% \textit{Incremental iterative processing} and related extensions we
% make to the core SciDB engine.  Section~\ref{sec:overlap} covers
% \textit{iterative overlap processing} and the additional
% \textit{mini-iteration optimization}. In Section~\ref{sec:multi-res},
% we describe the main idea around \textit{multi-resolution
%   optimization} and its advantage in case of changes in input
% data. Finally, Section~\ref{sec:eval} evaluates the proposed
% techniques on a real dataset from the astronomy domain.

%\magda{Need to say how this approach differs from related work
%and what non-trivial problem it solves. Continue to use
%the example throughout the text.}
%
%\magda{Once you write down your ideas once, you should go back and
%  think about how you need to tell the story in a way that will be
%  easy for readers to follow. They need to be convinced that (1) you
%  are solving an important problem, (2) you are solving a non-trivial
%  problem, (3) your solution improves on prior art, (4) your solution
%  is elegant and it works, and (5) why you chose this solution and not
%  another.}

\section{Motivating Applications}
\label{sec:examples}

We start by presenting three array-oriented, iterative
applications. We use these applications as examples throughout the
paper and also in the evaluation.

\begin{figure}[t]
\begin{center}
 \subfigure[Single image]{
\includegraphics[width=0.30\linewidth]{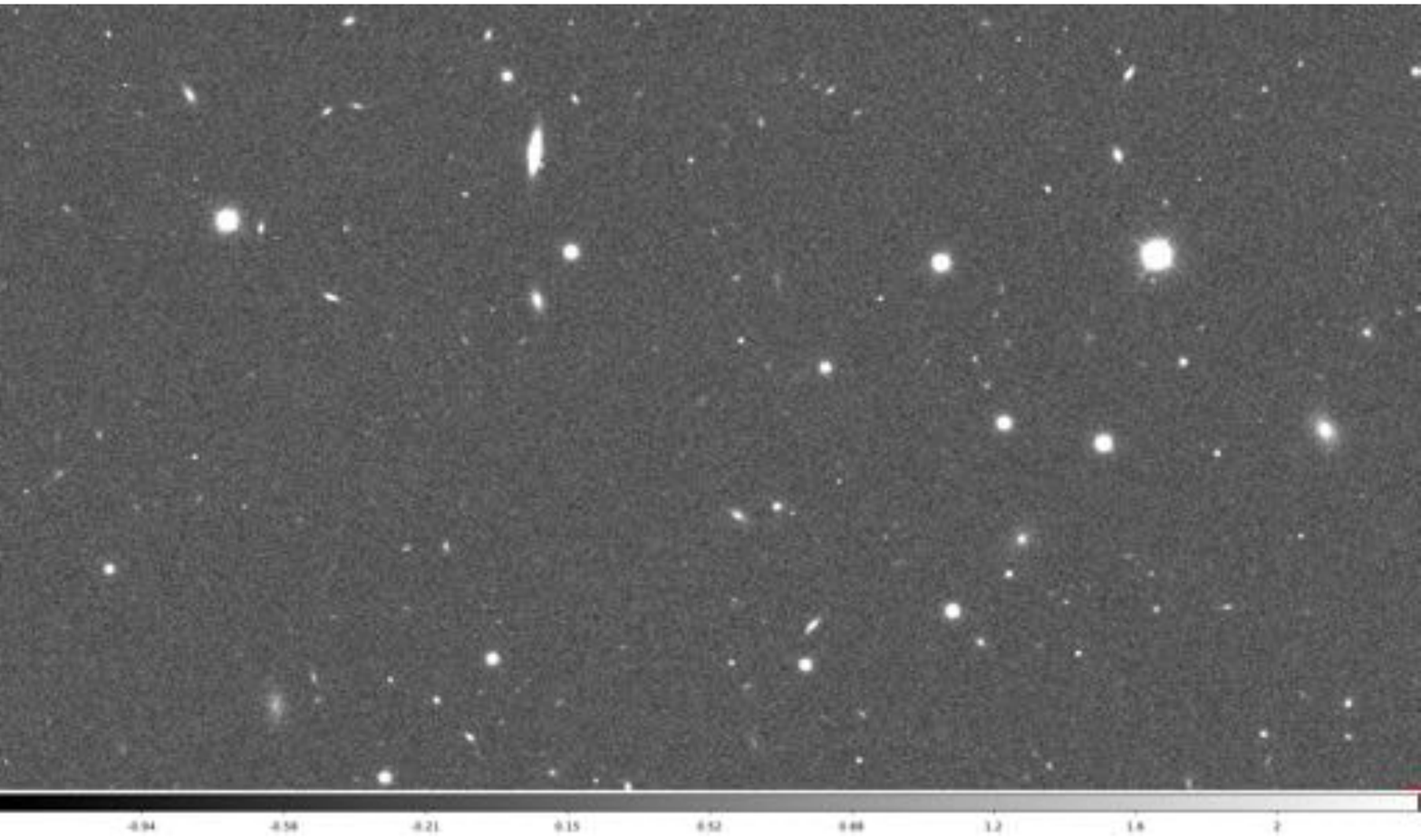}
 \label{fig:single}
 }
 \subfigure[\small{Co-added image}]{
\includegraphics[width=0.30\linewidth]{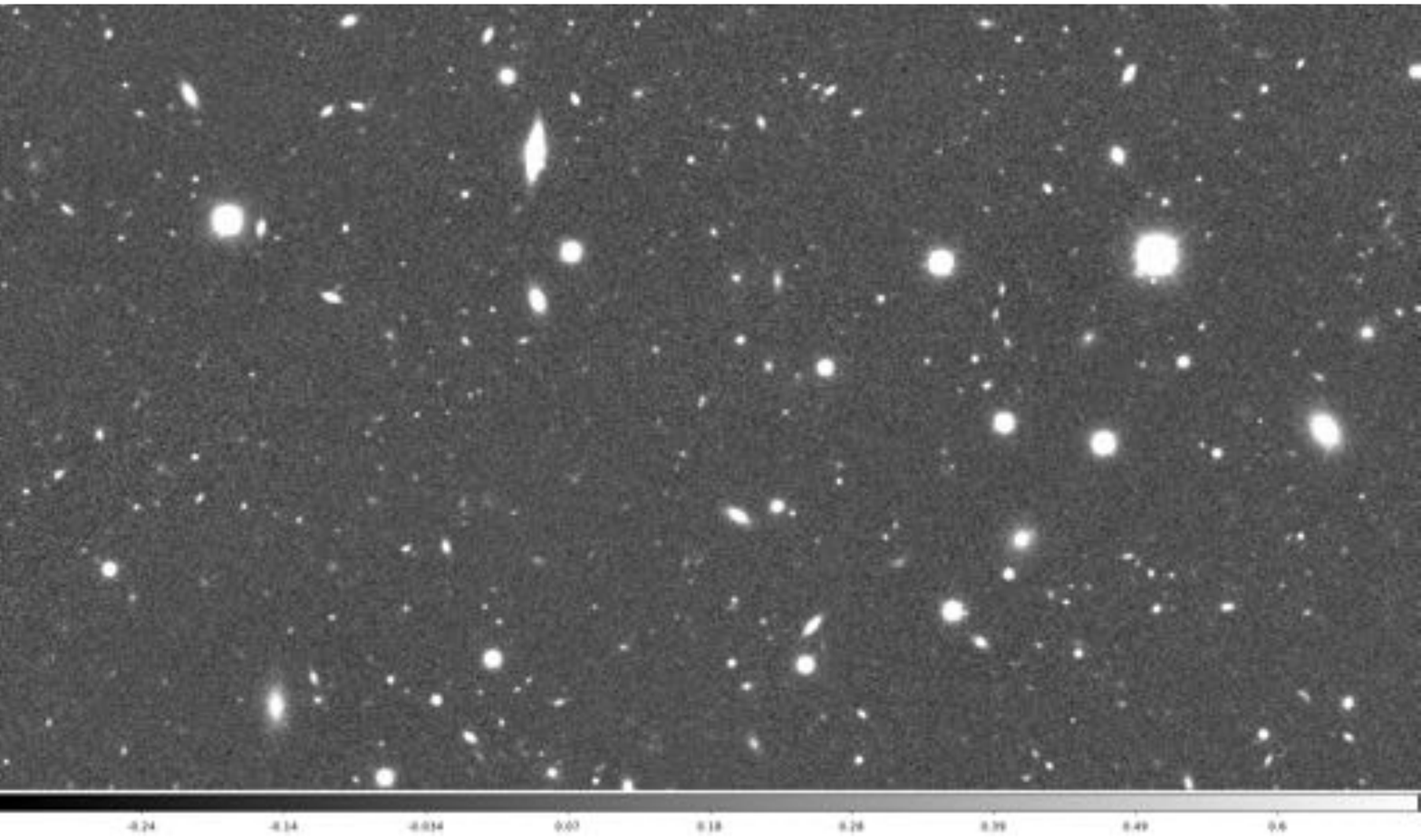}
 \label{fig:coadd}
 }
  \caption{Illustrative comparison of a \textit{single} telescope image and its corresponding \textit{co-added} image. 
   Many faint objects become visible after co-addition.}
 \label{fig:coadd-single}
\end{center}
\vspace{-10pt}
\end{figure}

\begin{example}
\label{ex:lsst}
\textbf{Sigma-clipping and co-addition in LSST images (SigmaClip):}~The
Large Synoptic Survey Telescope (LSST~\cite{lsst}) is a large-scale,
multi-organization initiative to build a new telescope and use it to
continuously survey the visible sky. The~\texttt{LSST} will generate
tens of TB of telescope images every night.  Before the telescope
produces its first images, astronomers are testing their data analysis
pipelines using realistic but simulated images.

%storage techniques, and data exploration using realistic
%but simulated images.

%The planned survey will
%cover more sky with more visits than any survey before. The novelty of
%he project means that no current dataset can exercise the full
%complexity of the data expected from the LSST. For this reason, before

When analyzing telescope images, some sources (a ``source" can be a
galaxy, a star, etc.)  are too faint to be detected in one image but
can be detected by stacking multiple images from the same location on
the sky. The pixel value (\texttt{flux} value) summation over all
images is called image
\textit{co-addition}. Figure~\ref{fig:coadd-single} shows a
\textit{single} image and the corresponding \textit{co-added} image.
Before the co-addition is applied, astronomers often run a
``sigma-clipping" noise-reduction algorithm. The analysis in this case
has two steps: (1) outlier filtering with ``sigma-clipping" and then
(2) image co-addition. Listing~\ref{code:sigma-pseudo} shows the
pseudocode for both steps. Sigma-clipping consists in grouping all
pixels by their (x,y) coordinates. For each location, the algorithms
computes the mean and standard deviation of the flux.  It then sets to
null all cell values that lie $k$ standard deviations away from the
mean.  The algorithm iterates by re-computing the mean and standard
deviation. The cleaning process terminates once no new cell values are
filtered out. Throughout the paper, we refer to this application as
\textbf{SigmaClip}. \marker

%Finally, those groups of cells often influence the value
%of some other cells in the next iteration by some pre-computed
%aggregated value.  \marker
\end{example}

\begin{example}
\label{ex:detection}
\textbf{Iterative source detection algorithm (SourceDetect):}~Once
telescope images have been cleaned and co-added, the next step is
typically to extract the actual sources from the images.

%\textit{Source detection} is a time consuming operation
% that should be
%parallelized by breaking down the large raw image into multiple
%maller images.  This algorithm is often implemented as a user-defined
%function that involves entire scan of the raw data followed by a
%grouping phase that groups observations into ones that represent the
%same object. The other way to implement this algorithm is iterative
%but avoids the overhead of grouping phase at the end.  A simplified
%version of the iterative source detection algorithm is described in

The pseudocode for a simple source detection algorithm is shown in
Listing~\ref{code:SourceDetect}. Each non-empty cell is initialized with a
unique label and is considered to be a different object. At each
iteration, each cell resets its label to the minimum label value
across its neighbors. Two cells are neighbors if they are adjacent.
This procedure continues until the algorithm converges. 
We refer to this application as \textbf{SourceDetect}.  \marker
\end{example}

%\emad{I am not sure I can do k-means evaluation on time ... we can still mention it as example though.}
\begin{example}
\label{ex:kmeans}
\textbf{K-means clustering algorithm (KMeans):}~ In many domains,
clustering algorithms are commonly used to identify patterns in
data. Their use extends to array data. We consider in particular
K-means clustering on a 2D array~\cite{astroMLText}. 
K-means clustering works as follows: It assigns each
cell randomly to one of the $k$ clusters. It computes the centroid of
each cluster.  It iterates by re-assigning each cell to its nearest
cluster. We refer to this application as \textbf{KMeans}. \marker
 %\magda{Need to check details of how Andy applies this
%technique for his array data.} 
\end{example}

These applications illustrate two important properties of iterative
computations over arrays.  First, the goal of an iterative computation
is to take an array from an initial state to a final state by
iteratively refining its content. The \texttt{SigmaClip} application,
for example, starts with an initial 3D array containing 2D images
taken at different times. Each iteration changes the cell values in
this array.  The iteration terminates when no cell changes across two
iterations. Second, the value of each cell at the next iteration is
determined by the values of a \textit{group} of cells with some common
\textit{characteristics} at the current iteration.  Those
characteristics are often mathematically described for any given cell
in the array. For \texttt{SigmaClip} those are ``\textit{all pixel 
values at the same (x,y) location}". Interestingly, unlike \texttt{SigmaClip}, 
where each group of cells at the same
$(x,y)$ location influences \textit{many} cell-values at the next
iteration, in the \texttt{SourceDetect} algorithm any given cell
$(x,y)$ is influenced by a \textit{unique} group of cells, which are
its adjacent neighbors. These groups of cells partially overlap with
each other, which complicates parallel processing as we discuss in
Section~\ref{sec:overlap}.

%In the rest of the paper we use
%examples~\ref{ex:lsst} and~\ref{ex:detection} to present the approach
%and evaluate it.

%A configurable $\epsilon$ threshold defines
%the range of neighborhood. This procedure continues until no change in
%the cell values occurs, or equivalently, when the iterative algorithm
%\textit{converges}. 

%bounded by $\epsilon$ distance from the $(x,y)$
%location and more interestingly each group is partially overlapped
%with many other groups. 

\begin{Listing}[t]
\scriptsize
\begin{verbatim}
Input: Array A with pixels from x-y images over time. 
//Part 1: Iterative sigma-clipping
While(some pixel changes in A)
 For each (x,y) location
  Compute mean/stddev of all pixel values at (x,y).
  Filter any pixel value that is k 
  standard deviations away from the mean  
//Part 2: Image co-addition
Sum all non-null pixel values grouped by x-y
\end{verbatim}
\caption{Pseudocode for SigmaClip application}
\label{code:sigma-pseudo}
\vspace{-10pt}
\end{Listing}

\begin{Listing}[t]
\scriptsize
\begin{verbatim}
Input: Co-added Array A with uniquely labeled pixels from 
       all the x-y images. 
Input: int r, the adjacency threshold. 
While(some pixel changes in A)
 For each (x,y) location
  Compute the minimum label of all pixel values (x',y')
  with x-r <= x'<= x+r  and  y-r <= y'<= y+r. 
  Update (x,y) with the minimum label. 
\end{verbatim}
\caption{Pseudocode for SourceDetect application}
\label{code:SourceDetect}
\vspace{-10pt}
\end{Listing}

%\begin{algorithm}[t] 
%\caption{Iterative \texttt{source-detection} function in lsst images.}
%\label{alg:sd-sql}
%\begin{algorithmic}[1]
%\Function{source-detection}{$A$,$\epsilon$}
%\State Input: Array $A$ $<$int $label$$>$[$x$,$y$] 
%\State Input: $\epsilon$ neighborhood threshold 
%\While{(some pixels $A[x,y]$ has changed)}
%\State $A[x,y]$ = select $min(A.label)$ from $A$ group by $[x + \epsilon][y + \epsilon]$ \Comment window sliding aggregation.
%\EndWhile
%\EndFunction
%\end{algorithmic}
%\end{algorithm}
\section{Iterative Array-Processing Model}
\label{sec:model}

We start with a formal definition of an array similar to Furtado and
Baumann~\cite{furtado:99}: Given a discrete coordinate set $S = S_1
\times \ldots \times S_d$, where each $S_i, i \in [1,d]$ is a finite
totally ordered discrete set, an array is defined by a d-dimensional
domain $D = [I_1, \ldots, I_d]$, where each $I_i$ is a subinterval of
the corresponding $S_i$.  Each combination of dimension values in D
defines a \textit{cell}. All cells in a given array $A$ have the same
type $T$, which is a tuple. $cells(A)$ is the set of all the cells in
array $A$ and function $V: cells(A) \rightarrow T$ maps each cell in
array $A$ to its corresponding tuple with type $T$. In the rest of the
paper, we refer to the dimension $x$ in array $A$ as $A[x]$ and to
each attribute $y$ in the array $A$ as $A.y$.

\begin{figure}[t]
\includegraphics[width=0.80\linewidth]{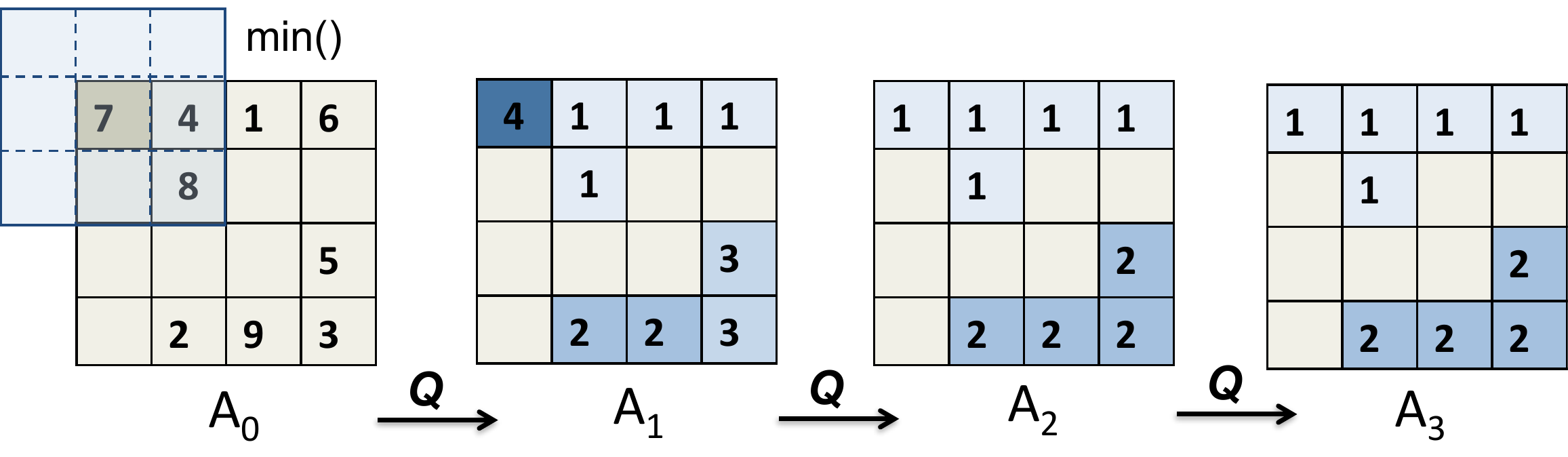}
 \caption{Iterative array $A$ and its state at each iteration for the \texttt{SourceDetect} application. $\{Q^{f^{\pi},\delta^{\pi}}_{cells(A)}: \forall c_{i,j} \in cells(A) \; i \in I_1 \;\&\; j \in I_2 \; \}$ where $I_1=I_2=\{1,2,3,4\}$ are the sets of dimension values, $f^{\pi}$ applies $min()$ aggregate on each group of cells, $\delta^{\pi}$ simply stores the aggregated value in each cell $c_{i,j}$, and $\pi: (x,y) \rightarrow [x \pm1][y \pm1]$. At each iteration, a sliding window scans through all the cells.}
\label{fig:iter-example-2}
\vspace{-10pt}
\end{figure}

In SciDB, users operate on arrays by issuing declarative queries using
either the Array Query Language (AQL) or the Array Functional Language
(AFL). The \texttt{select} statements in
Algorithm~\ref{alg:sigma-clip} in Section~\ref{sec:inc} are examples
AQL queries.  AQL and AFL queries are translated into query plans in
the form of trees of array operators. Each operator $O$ takes one or
more arrays as input and outputs an array: $O: A \rightarrow A$ or $O:
A \times A \rightarrow A$.

In an iterative computation, the goal is to start with an initial
array $A$ and transform it through a series of operations in an
iterative fashion until a termination condition is satisfied. The
iterative computation on $A$ typically involves other arrays,
including arrays that capture various intermediate results (\eg,
arrays containing the average and standard deviation for each $(x,y)$
location in the \texttt{SigmaClip} application) and arrays with
constant values (\eg, a connectivity matrix in a graph application).

%The examples are the intermediate array that captures 
%mean and standard deviation for each $(x,y)$ location in \texttt{SigmaClip} application
%nd the constant array that captures the edge structure of the cells in graph algorithms.

One can use the basic array model to express iterative
computations. The body of the loop can simply take the form of a
series of AQL or AFL queries.  Similarly, the termination condition
can be an AQL or AFL query. In Section~\ref{sec:inc}, the first
function in Algorithm~\ref{alg:sigma-clip} illustrates this approach.

To enable optimizations, however, we extend the basic array model with
constructs that capture in greater details how iterative applications
process arrays. We start with some definitions.

\begin{definition}
  We call an array \textit{iterative} if its cell-values are updated
  during the course of an iterative computation. The array starts with
  an initial state $A_0$. As the iterative computation progresses, the
  array goes through a set of states $A_1$, $A_2$, $A_3$, $\ldots$,
  until a final state $A_N$. Note that all $A_i$ have the same schema. In other words, the shape of an iterative array does not change.  
\end{definition}

Figure~\ref{fig:iter-example-2} shows a (4$\times$4) iterative array
that represents a tiny telescope image in the \texttt{SourceDetect}
application. In the initial state, $A_0$, each pixel with a flux value
above a threshold is assigned a unique value. As the iterative
computation progresses, adjacent pixels are re-labeled as they are
found to belong to the same source. In the final state $A_3$, each set
of pixels with the same label corresponds to one detected source.

%and the state of that array at each iteration.  
%$A_i$ represents state
%of the iterative array $A$ at iteration $i$. $A_0$ is the original
%iterative array (we continue to use the figure throughout the
%section).

Iterative applications typically define a termination condition that
examines the cell-values of the iterative array: 

\begin{definition}
  An iterative array $A$ has \textit{converged}, whenever
  $T(A_i,A_{i+1}) \le \epsilon$ for some aggregate function $T$. $T$
  is the \textit{termination} condition. $\epsilon$ is a
  user-specified constant. 
\end{definition}

In Figure~\ref{fig:iter-example-2}, convergence occurs at
iteration 3 when $\epsilon=0$ and the termination condition $T$ is
the count of differences between $A_i$ and $A_{i+1}$. Our ArrayLoop 
system represents $T$ as AQL function. 

%\magda{Maybe here state the termination condition for each
%of our three applications.}

An iterative array computation takes an iterative array, $A$, and applies to
it a computation $Q$ until convergence:

\begin{align}
\label{eq:iterative-computation}
A_{0} \xrightarrow{Q} A_{1} \xrightarrow{Q} \dots \xrightarrow{Q} A_i \xrightarrow{Q} A_{i+1}
\end{align}

%Figure~\ref{fig:iter-example-2} is an iterative computation with three steps that represents ``iterative source detection" algorithm. 

%\magda{We need to define what is a valid $Q$: Can we handle any
%  sequence of AQL queries as Q?}
\noindent where $Q$ is a sequence of valid AQL or AFL queries. 
At each step, $Q$ can either update the entire array or only some
subset of the array. We capture the distinction with the notion of
\textit{major} and \textit{minor} iteration steps:

%non-empty cells in $A$ at the same time. Otherwise it is a
\begin{definition}
  A state transition, $A_i \xrightarrow{Q} A_{i+1}$, is a
  \textit{major step} if the function $Q$ operates on all the 
  cells in $A$ at the same time. Otherwise it is a
  \textit{minor step}.
\end{definition}

%\textit{associative} and \textit{commutative} 

The array state $A_{i,j}$ represents the state of the iterative array
after $i$ major steps followed by $j$ minor steps.  We are interested
in modeling computations where each major step can be decomposed into
a set of minor steps that can be evaluated in parallel.  That is, a
major step $Q_i$ can be expressed as a set of minor steps $q_i$ such
that $\sigma$, $Q_i = q_{i,\sigma_1} \cdot q_{i,\sigma_2} \dots
q_{i,\sigma_{n-1}} \cdot q_{i,\sigma_n}$.

%A set of minor steps represents a major step if those minor steps cover all cells in
%the iterative array. That is, if together they apply $Q$ to all cells
%of the array, possibly with replication. Q can always be executed using
%either minor steps or major steps.

The iterative array computation in Equation~\ref{eq:major-mini-ex}
includes $(i+1)$ major steps. The first line illustrates the
transition of iterative array $A$ in major steps and the second line
illustrates the possible minor steps that can replace the major step
$Q_{i+1}$.  A termination condition check always occurs between two
states of an iterative array after a major step.

\begin{align}
\label{eq:major-mini-ex}
 A_{0} \xrightarrow{Q_{1}} A_{1} \xrightarrow{Q_{2}}  \dots A_{i-1} \xrightarrow{Q_{i}} A_{i} \xrightarrow{Q_{i+1}} A_{i+1}  \\
\overbrace{A_{i} \xrightarrow{q_{i,1} \cdot q_{i,2} \dots q_{i,j-1} \cdot q_{i,j}}  A_{i+1}} \nonumber
\end{align}

\begin{figure}[t]
\includegraphics[width=0.75\linewidth]{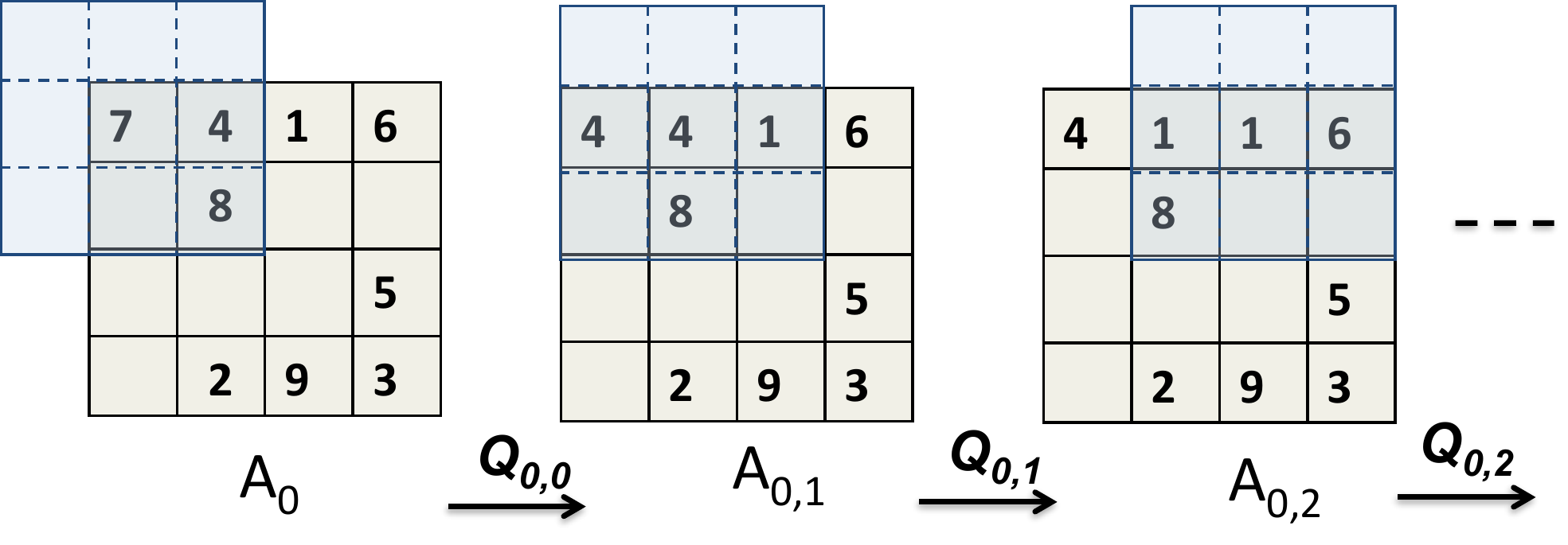}
\caption[Iterative array A and its state after three minor steps.]{Iterative array $A$ and its state after three minor steps,
  each of the form: $Q_{i,j}=Q^{f^{\pi},\delta^{\pi}}_{c_{i,j}}$ where $c_{i,j}$ is the cell at $A[i][j]$,
  $f^{\pi}$ applies $min()$ aggregate, $\delta$ simply stores the aggregate result as the new value in cell
  $c_{i,j}$, and $\pi: (x,y) \rightarrow [x \pm1][y \pm1]$}
 \label{fig:iter-example-1}
 \vspace{-10pt}
\end{figure}

Figure~\ref{fig:iter-example-2} shows an iterative array computation
with only major steps involved, while Figure~\ref{fig:iter-example-1}
presents the same application but executed with minor steps.

%$A_i$ represents the state of the iterative array
%$A$ at major step \magda{what is a major step?} $i$ where $A_0$ is the
%original array and $A_k \; k \ge 0 $ is the iterative array $A$ where
%the iterative function is applied at least $k$ times to all the cells
%in $A$. \magda{Again, it would help here to refer back to an example
%  introduced in Section 2.} . \magda{Again,
%  refer back to example to illustrate the notion of convergence and
%  termination function T.}

%\magda{You may want to call-out the definitions better. Look at Section 3.3
%in this paper for example: \url{http://homes.cs.washington.edu/~magda/papers/khoussainova-vldb12.pdf}}

%An ideal iterative model is the one that abstracts away different
%iterative array computation strategies and mainly covers the semantics
%of the iterative algorithm. To complete our array iterative model
%discussion, we have to provide the definition of the \textit{array
%  iterative function} $Q$.  

We further observe from the example applications in
Section~\ref{sec:examples} that the functions $Q$ often
follow a similar pattern.

First, the value of each cell in iterative array $A_{i+1}$ that is
updated by $Q$ only depends on values in \textit{nearby cells} in
array $A_{i}$. We capture this spatial constraint with a function
$\pi$ that specifies the mapping from output cells back onto input
cells:

\begin{definition}

\label{def:pi}
%\magda{How is this different from a regrid? SciDB's regrid
%does not seem to allow overlapping windows but that seems like
%it's an artificial constraint. I think that we should say that
%Q is a type of regrid operation.}\emad{in case of the k-means the assignment function is not a regrid because 
%the common characteristics of the grouping cells is their attribute value ... if we omit this use-case then $\pi$ is an overlapping regrid.}
$\pi$ is an \textit{assignment} function defined as $\pi:
cells(A) \rightarrow \mathcal P \left(cells(A)\right)$, where
$cells(A)$ is the set of all the cells in array $A$ and $\mathcal P()$
is the powerset function.  
%$\pi$ assignment function identifies the
%set of cells in iterative array $A$ that contribute to given cell in
%the next step of the
%teration. %s a pair (D,T) where $D$ is a list of index values of $A$ and $T$ is the type of the cell contents in $A$.
\end{definition}

\begin{figure}[t]
\begin{center}
 \subfigure[\texttt{SourceDetect}]{
\includegraphics[width=0.25\linewidth]{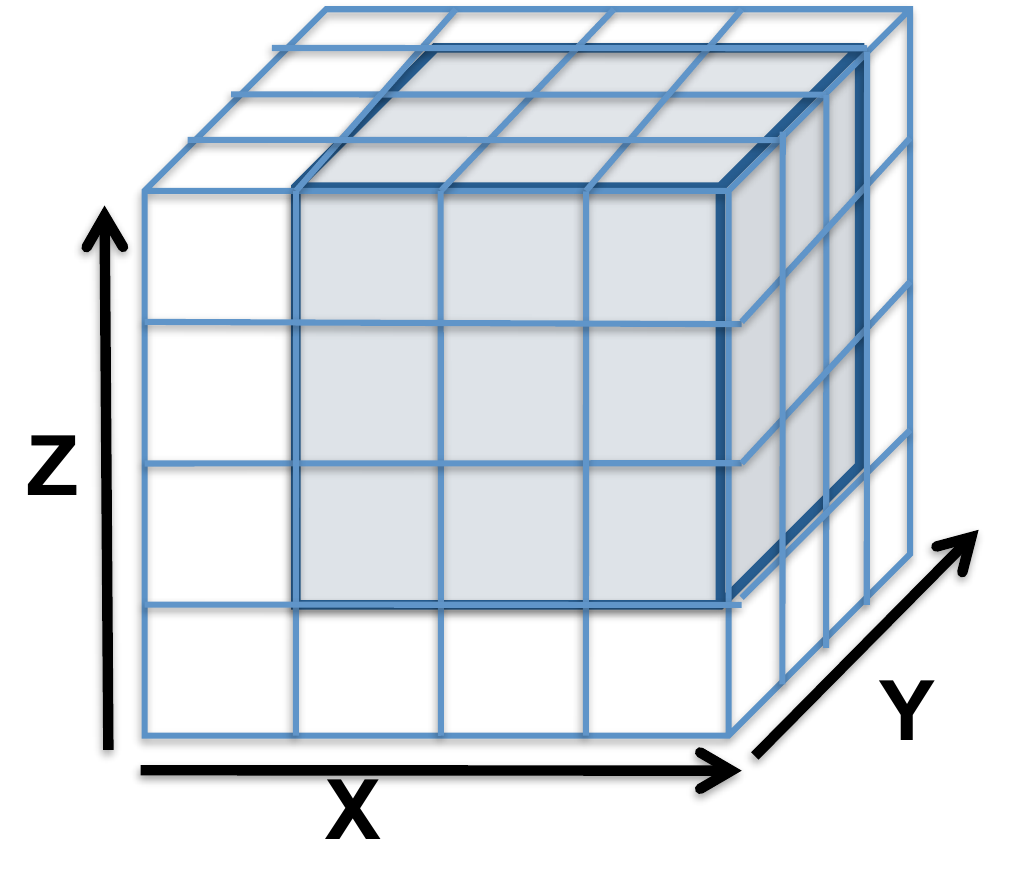}
 \label{fig:w1}
 }
 \hspace{30pt}
 \subfigure[\texttt{SigmaClip}]{
\includegraphics[width=0.25\linewidth]{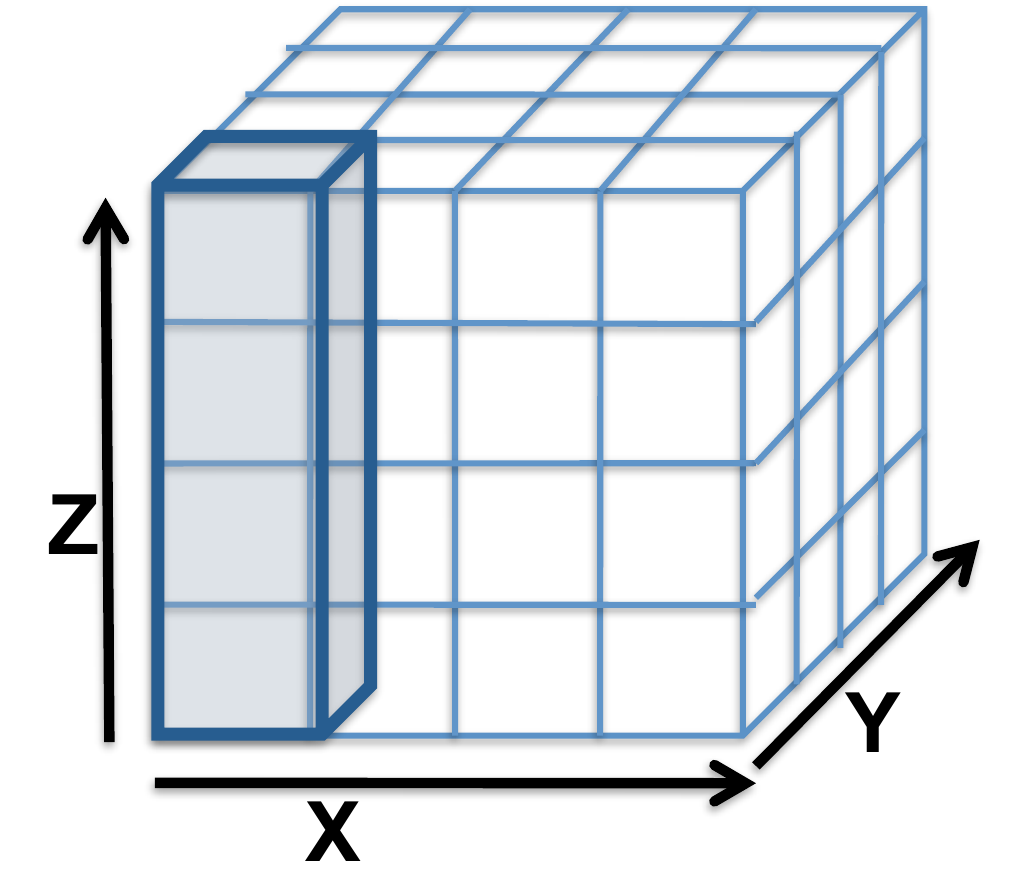}
 \label{fig:w2}
 }
\caption[Two examples of window assignment functions.]{Two examples of window assignment functions: (a) $\pi_1: (x,y,z) \rightarrow [x\pm1][y\pm1][z\pm1]$, the associated window is highlighted for the cell at $(2,1,2)$. (b) $\pi_2: (x,y,z) \rightarrow [x][y]$, the associated window is highlighted for all the cells at $(x,y,z)$ with $z=0$.}
\label{fig:window}
\end{center}
\vspace{-10pt}
\end{figure}

Figure~\ref{fig:window} illustrates two examples of assignment
functions. Our ArrayLoop system supports two types of assignment
functions: \textit{windowed} functions such as those illustrated in
Figure~\ref{fig:window} and \textit{attribute} assignment
function. The latter occur in applications such as K-means clustering
described in Example~\ref{ex:kmeans}: $\pi: (x,y) \rightarrow label$
where all the cells with the same label are grouped together.

%$\mathbb{R}_1,\dots,\mathbb{R}_k$
\begin{definition}
$f^{\pi}$ is an aggregate function defined as $f^{\pi}: cells(A) \rightarrow \tau$. $f^{\pi}$ groups the cells of the array $A$ according
to assignment function $\pi$, with one group of cells per cell in the array $A$. It
then computes the aggregate functions separately for each group. The aggregate result is stored in tuple $\tau$.

\end{definition}

Finally, $Q$ updates the output array with the computed aggregate values:

\begin{definition}
  $\delta^{\pi}: (cells(A),f^{\pi})
  \rightarrow cells(A)$ is a \textit{cell-update} function. It updates
  each cell of the array $A$ with the corresponding tuple $\tau$
  computed by $f^{\pi}$ and the current value of the cell itself.

%which is
%  defined as $\delta^{\pi}(c,f^{\pi})$ for some cells $c \in cells(A)$
%  which is identified by $\pi$ function.
\end{definition}

%\magda{We need to argue that the above capture many types of iterative applications.
%The way to do this is to show how each application from Section 2 can be defined in terms
%of the above concepts. }
%
%\magda{We may want to give the above functions names, such as
%  ``Stencil definition'' and ``Cell update'' functions.}\emad{I named them assignment, aggregation and cell-update functions.}
  
These three pieces together define the iterative array computation $Q^{f^{\pi},\delta^{\pi}}_C$ as follows:

\begin{definition}
  \textit{An iterative array computation} $Q^{f^{\pi},\delta^{\pi}}_C$ on the subset of cells
  $C$ where $C \in \mathcal P(cells(A))$ generates subset of cells $C^{'} \in \mathcal P(cells(A))$ such that $\forall c \in C$ and $c^{'} \in C^{'}$
  $c^{'} = \delta^{\pi}(c,f^{\pi}(c))$ where $c$ and $c^{'}$ are two corresponding cells in those subsets.  
%  is defined as a pair
%  $(f^{\pi},\delta^{\pi}(c_i,f^{\pi}))$  in the iterative array $A$, where the assignment function
%  $f^{\pi}(c_i)$ computes an aggregate value of all the cells that
%  influence cell $c_i$ at the next iteration and the
%  $\delta^{\pi}(c_i,f^{\pi})$ update-function changes $c_i$ based on
%  the aggregate value and the current value of $c_i$.
%\begin{equation}
%\label{eq:Q}
%Q=\{(f^{\pi}(c_i),\delta^{\pi}(c_i,f^{\pi})): \forall c_i \in C\}
%\end{equation}
%where $c_i  \in C$ is a subset of cells in the iterative array. 
\end{definition}

In the example from Figures~\ref{fig:iter-example-2}
and~\ref{fig:iter-example-1}, which illustrate the
\texttt{SourceDetect} application, the goal is to detect all the
clusters in the array $A$, where each cell $p_1=(x_1,y_1)$ in a
cluster has at least one neighbor $p_2=(x_2,y_2)$ in the same cluster
such that $|x_1 - x_2| \le 1$ and $|y_1 - y_2| \le 1 $, if it is not a
single-cell cluster.  In this application, $\pi$ is the 3X3 window
around a cell. We slide the window over the array cells in major
order. At each \textit{minor} step, at each cell $c_{i,j}$ at the
center of the window, we apply an iterative array computation
$Q_{i,j}=Q^{f^{\pi},\delta^{\pi}}_{c_{i,j}}$ where $f^{\pi}$ applies a
$min()$ aggregate over the 3x3 window, $\pi$, and $\delta^{\pi}$ is a
cell-update function that simply stores the result of the $min()$
aggregate into cell $c_{i,j}$.  Figure~\ref{fig:iter-example-1}
illustrates three steps of this computation. Notice that the output of
the iterative array computation $Q_{0,0}$ becomes the input for
$Q_{0,1}$ and so on. Another strategy is to have many windows grouped
and applied together.  In other words, instead of applying the iterative
array computation per cell, we apply $Q^{f^{\pi},\delta^{\pi}}_{C}$ on
a group of cells $C \in \mathcal P \left(cells(A)\right)$ in one
\textit{major} step. Note that when using minor steps, the output of
each minor step serves as input to the next step.  In contrast, when
using major steps, the iterative array computations see the original
array state at the beginning of that
iteration. Figure~\ref{fig:iter-example-2} shows the iterative array
computation for the latter strategy. The former strategy has less
expensive steps than the latter strategy, but it requires more steps
to converge.

In our model, we encapsulate all the elements of the model in a \textit{FixPoint} operator: 
\begin{align}
\label{eq:Fixpoint}
&FixPoint(A,\pi,f,\delta,T,\epsilon) 
\end{align}

%\subsection{ArrayLoop optimizer}

With our model, the user specifies the logic of the iterative
algorithm without worrying about the way it is going to be
executed. Our model can be implemented and executed on top of various
array execution engines. In the rest of the paper, we describe how the
queries specified in our model are rewritten and efficiently run in
the SciDB array engine. The execution strategy in SciDB uses only
major steps.  Mini-step iterations, i.e. asynchronous execution, is
left for future study.

%\begin{algorithm}[t]
%\scriptsize
%\caption{Iterative algorithm, na\"{\i}ve approach}
%\begin{algorithmic}[1]
%\While{$A_i \ne A_{i+1}$}
%\State run AQL $Q_1$
%\State $\dots$
%\State run AQL $Q_k$
%\EndWhile
%\end{algorithmic}
%\label{iter:naive}
%\end{algorithm}
%(shown in Algorithm~\ref{iter:naive})

\section{Iterative Array Processing}
\label{sec:exec}

%To execute an iterative array computation in SciDB, a na\"{\i}ve
%approach is to simply iteratively invoke array queries from a
%high-level scripting language such as SciDB-Py~\cite{scidb-py}, a
%python interface for SciDB. This approach, however, prevents or at
%least complicates the automated optimizations of iterative
%computations. Instead, 

We extend SciDB-Py~\cite{scidb-py} with a python $FixPoint()$ operator
following the model from Section~\ref{sec:model}.  We also develop an
optimizer module that we name ArrayLoop.  The user encapsulates its
iterative algorithm in the $FixPoint()$ operator. The ArrayLoop
optimizer sits on top of SciDB. ArrayLoop rewrites a $FixPoint(A, \pi, f, \delta, T, \epsilon)$ operator into the AQL queries in Listing~\ref{code:fixpoint-translate} that it wraps with an internal while loop. 

\texttt{is\_window} helper function in Listing~\ref{code:fixpoint-translate} clarifies whether the window assignment function translates to a window aggregate or a group-by aggregate. ArrayLoop translates a window assignment function to a group-by aggregate if mapping is from a set of input dimensions to one of its subsets. If mapping is from a set of dimensions to the same set of dimensions with additional offsets per dimension, then ArrayLoop translates it to window-aggregate. Supporting window assignment function that is a combination of group-by aggregate and window aggregate is left for future work. 

In addition, ArrayLoop also implements a set of query re-writing tasks in order to
leverage a series of optimizations that we develop:
\textit{incremental iterative processing}, \textit{overlap iterative
  processing}, and \textit{multi-resolution iterative processing}.

\begin{Listing}[t]
\scriptsize
\begin{verbatim}
Input:
  FixPoint(A,pi,f,delta,T,epsilon)
Output:  
  While (T(A,A_prev)  < epsilon)
   // Termination function T is also AQL function.
   // Compute the new aggregates from the current iterative array. 
   If (is_window(pi))
      G = SELECT f FROM A WINDOW PARTITIONED BY pi
   else
      G = SELECT f FROM A GROUP BY pi 
   // Combine the new aggregate with the old value of the cell. 
   S = SELECT * FROM G JOIN A ON <matching dimensions> 
   A_new = SELECT delta(S) FROM S 
   A_prev = A 
   A = A_new 
\end{verbatim}
\caption{Pseudocode for rewriting FixPoint operator}
\label{code:fixpoint-translate}
\vspace{-10pt}
\end{Listing}

ArrayLoop acts as a pre-processing module before executing the
iterative query in SciDB. Currently the majority of the ArrayLoop
implementation is outside the core SciDB engine. As future work, we
are planning to push the ArrayLoop python prototype into the core
SciDB engine. ArrayLoop relies on SciDB for features such as
distributed query processing, fault-tolerance, data distribution, and
load balancing. In the following sections, we describe each of the
three optimizations in more detail.

\section{Incremental Iterations}
\label{sec:inc}

%\magda{The contributions are buried in the text. We need to call them
%  out. I would recommend the following. Start by showing that manual
%  incremental processing is painful the way you do it. Then say that
%  we can automate it and hide the details. Before saying anyting else,
%  say that we need two things: (1) the ability to automatically
%  rewrite queries into incremental ones and (2) this special merge
%  operator. Only after giving the high-level idea, explain how we do
%  (1) and (2). I think we currently only explain (2). We need to
%  explain (1) also. Use Algorithm 4sec
%  early to show all that. In a separate subsection, show how we can
%  further optimize the approach by pushing deltas into the storage
%  layer. Again, start with the high-level idea and contribution and
%  then dive into the details. }
%
\label{sec:inc}
\begin{algorithm}[t] 
\tiny
\caption{\small \texttt{SigmaClip} application followed by image co-addition}
\label{alg:sigma-clip}
\begin{algorithmic}[1]
\renewcommand{\alglinenumber}[1]{\tiny #1.}
\Function{\textbf{sigma-clipping}}{$A$,$k$} \Comment{Na\"{\i}ve sigma-clip}
\State Input: Iterative Array $A$ $<$float $d$$>$[$x$,$y$,$t$] 
\State Input: $k$ a constant parameter.
\While{(some pixels $A[x,y,t]$ are filtered)}
\State  \colorbox[gray]{0.90}{\parbox[t]{\dimexpr\linewidth-\algorithmicindent}{$T[x,y]$ = select $avg(d)$ as $\mu$,  $stdv(d)$ as $\sigma$ from $A$ group by $x,y$\strut}}
\State \parbox[t]{\dimexpr\linewidth-\algorithmicindent}{$S[x,y,t]$ = select * from $T$ join $A$ on $T.x$ = $A.x \;and\; T.y$=$A.y$\strut}
\State  \parbox[t]{\dimexpr\linewidth-\algorithmicindent}{$A[x,y,t]$ = select $d$ from $S$ where $\mu - k \times \sigma \le d \le \mu + k \times \sigma$\strut}
\EndWhile
\EndFunction 
\Statex
\Function{\textbf{incr-sigma-clipping}}{$A$,$k$} \Comment{Incremental sigma-clip}
%\magda{(1)This algorithm is tedious to read. We should add an English-language comment before
%each line inside the iteration. (2) This algorithm is long and it is not our contribution. So
%it is hard to justify using all that space for this algorithm here. On the other hand,
%it is good to show the red boxes and how hard it is to write an incremenal query. So maybe it's ok
%to use that much space.}
\State Input: Array $A$ $<$float $d$$>$[$x$,$y$,$t$]. 
\State Input: $k$: a constant parameter.
\State Local: Array $C$ $<$int $c$,float $s$,float $s^2$$>$[$x$,$y$].
\State Local: $Collect \leftarrow \phi$ \Comment{Collects all the filtered points.}
\State Local: $Remain \leftarrow A$ \Comment{Keeps track of remaining points.}
\State $\Delta A \leftarrow A$ 
\While{($\Delta A$ is not empty)}
\State  \colorbox[gray]{0.90}{\parbox[t]{\dimexpr\linewidth-\algorithmicindent}{$T_1[x,y] \leftarrow $ select count$(d)$ as $c$,  sum($d$) as $s$, sum($d^2$) as $s^2$ from $\Delta A$ group by $x,y$\strut}}  \label{line:incr1} 
%\State \fcolorbox{red}{white}{$C_{prev} \leftarrow C$}\label{line:storage1}
\If{(first iteration)}
\State $C \leftarrow T_1[x,y]$ 
\Else 
\State \colorbox[gray]{0.75}{\parbox[t]{\dimexpr\linewidth-\algorithmicindent}{$\Delta$C[x,y] $\leftarrow$ select $C.c - T_1.c$ as $c$ , $C.s - T_1.s$ as $s$ , $C.s^2 - T_1.s^2$ as $s^2$ from $C$ join $T_1$ on $T_1.x$ = $C.x \;\&\; T_1.y$=$C.y$}\strut} \label{line:merge1} 
\EndIf
%\State \fcolorbox{red}{white}{\parbox[t]{\dimexpr\linewidth-\algorithmicindent}{$\Delta C \leftarrow$ select $C.c$, $C.s$, $C.s^2$ from $C$ join $C_{prev}$ on $C_{prev}.x$ = $C.x \;\&\; %C_{prev}.y$=$C.y$ where $C.c \ne C_{prev}.c$}\strut}\label{line:storage2}
\State  \colorbox[gray]{0.90}{\parbox[t]{\dimexpr\linewidth-\algorithmicindent}{$T[x,y] \leftarrow$ select $\frac{C.s}{C.c}$ AS $\mu$, $\sqrt[2]{\frac{C.s^2}{C.c} - (\frac{C.s}{C.c})^2}$ AS $\sigma$ from $\Delta C$\strut}}\label{line:incr2}
%\State  \colorbox[gray]{0.75}{merge$(A,T,T.\mu - k \times T.\sigma \le A.d \le T.\mu + k \times T.\sigma :A?\phi)$}
%\State  \fcolorbox{red}{white}{$A_{prev} \leftarrow A$}\label{line:storage3}
\State  \colorbox[gray]{0.75}{\parbox[t]{\dimexpr\linewidth-\algorithmicindent}{$S[x,y,t] \leftarrow$ select $A.d$, $T.\mu$, $T.\sigma$ from $T$ join $Remain$ on $T.x$ = $A.x \;\ and \; T.y$=$A.y$\strut}}\label{line:merge2} 
\State  \colorbox[gray]{0.75}{\parbox[t]{\dimexpr\linewidth-\algorithmicindent}{$\Delta A \leftarrow$ select $d$ from $S$ where $d \le \mu - k \times \sigma \; or \;  d \ge \mu + k \times\sigma$\strut}} \label{line:merge3} 
%\State \fcolorbox{red}{white}{\parbox[t]{\dimexpr\linewidth-\algorithmicindent}{$\Delta A \leftarrow$ select $A_{prev}.d$ from $A$ join $A_{prev}$ on $A_{prev}.x$ = $A.x \;\&\; A_{prev}.y$=$A.y \;\&\; A_{prev}.t$=$A.t$ where $A.d \ne A_{prev}.d$}\strut}\label{line:storage4}
\State \fcolorbox{red}{white}{$Remain\leftarrow \pi_{d}(S)$-$\Delta A$}\label{line:storage3} \Comment{Updates Remain.}
\State \fcolorbox{red}{white}{$Collect \leftarrow$ $\Delta A$}\label{line:storage4} \Comment{Adds the filtered points to Collect.}
\EndWhile
\State \fcolorbox{red}{white}{$A \leftarrow A$-$Collect$}\label{line:storage5}\Comment{Produces the final state for A.}
\EndFunction
\Statex
\Statex \textbf{co-addition phase:}
\State $R[x,y] \leftarrow$ select sum($A.d$) as $coadd$ from $A$ group by $x,y$
\end{algorithmic}
\end{algorithm}

\begin{figure}[t]
\begin{center}
 \subfigure[]{
\includegraphics[width=0.20\linewidth]{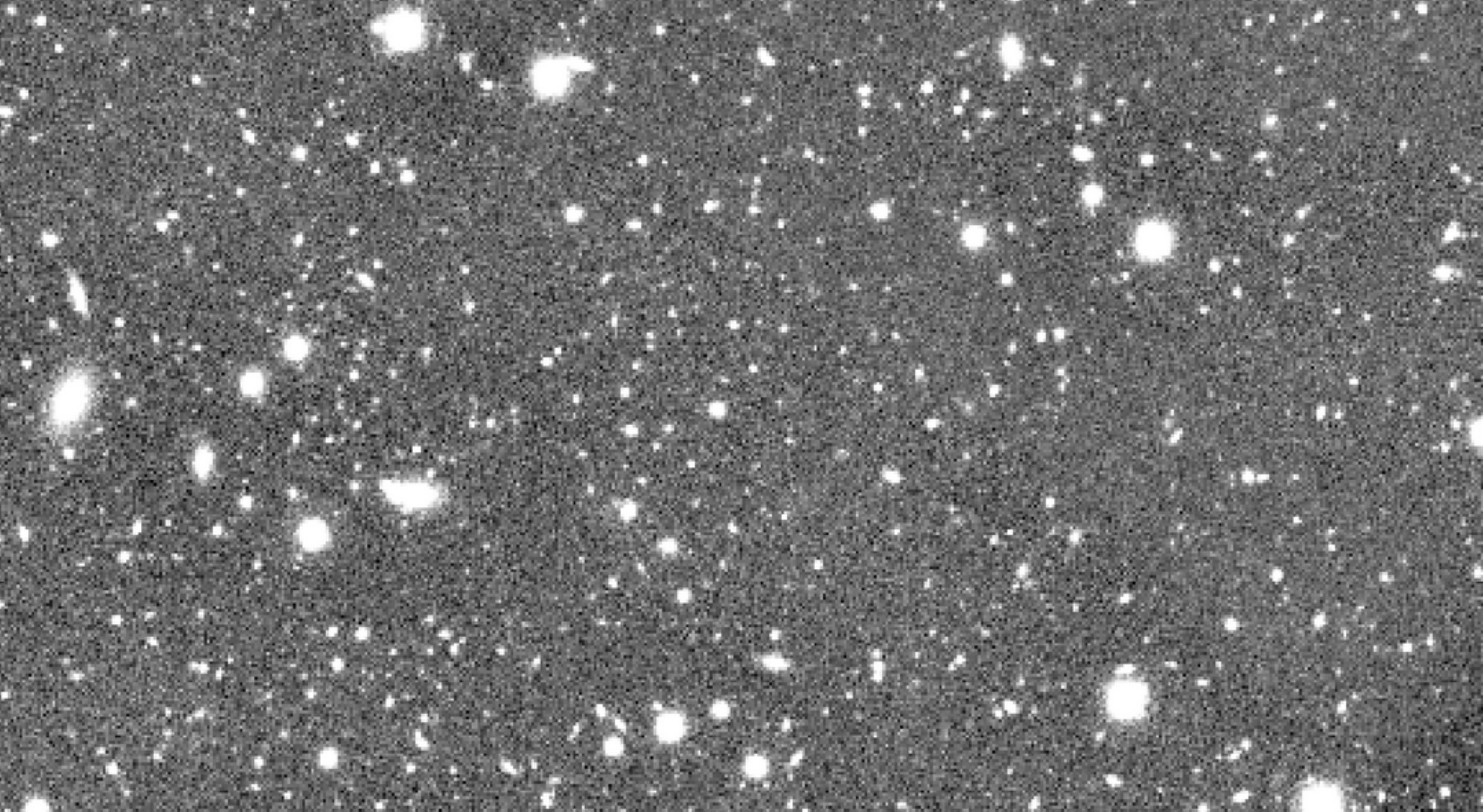}
 \label{fig:incr-1}
 }
 \subfigure[]{
\includegraphics[width=0.20\linewidth]{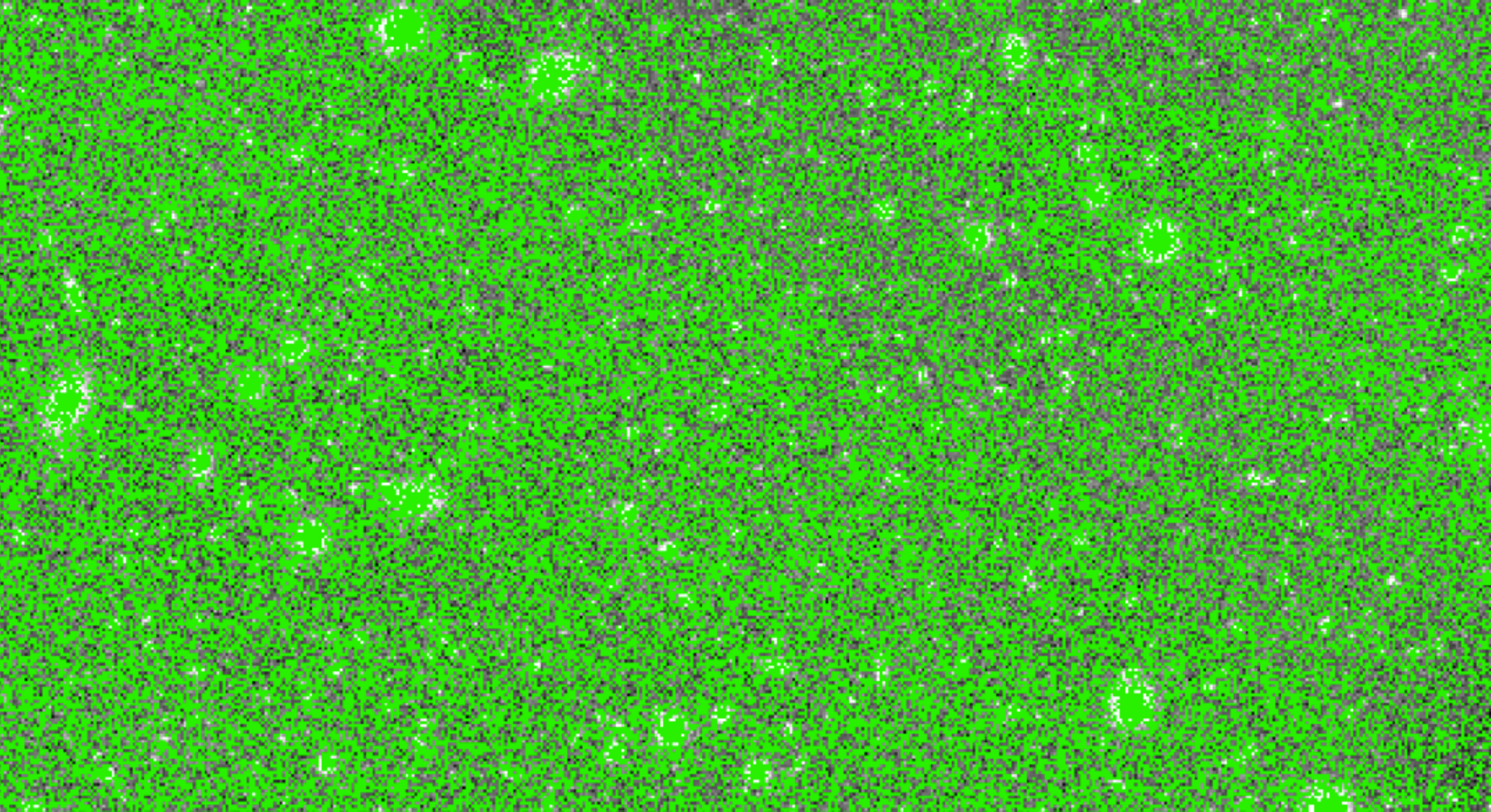}
 \label{fig:incr-2}
 }
 \subfigure[]{
\includegraphics[width=0.20\linewidth]{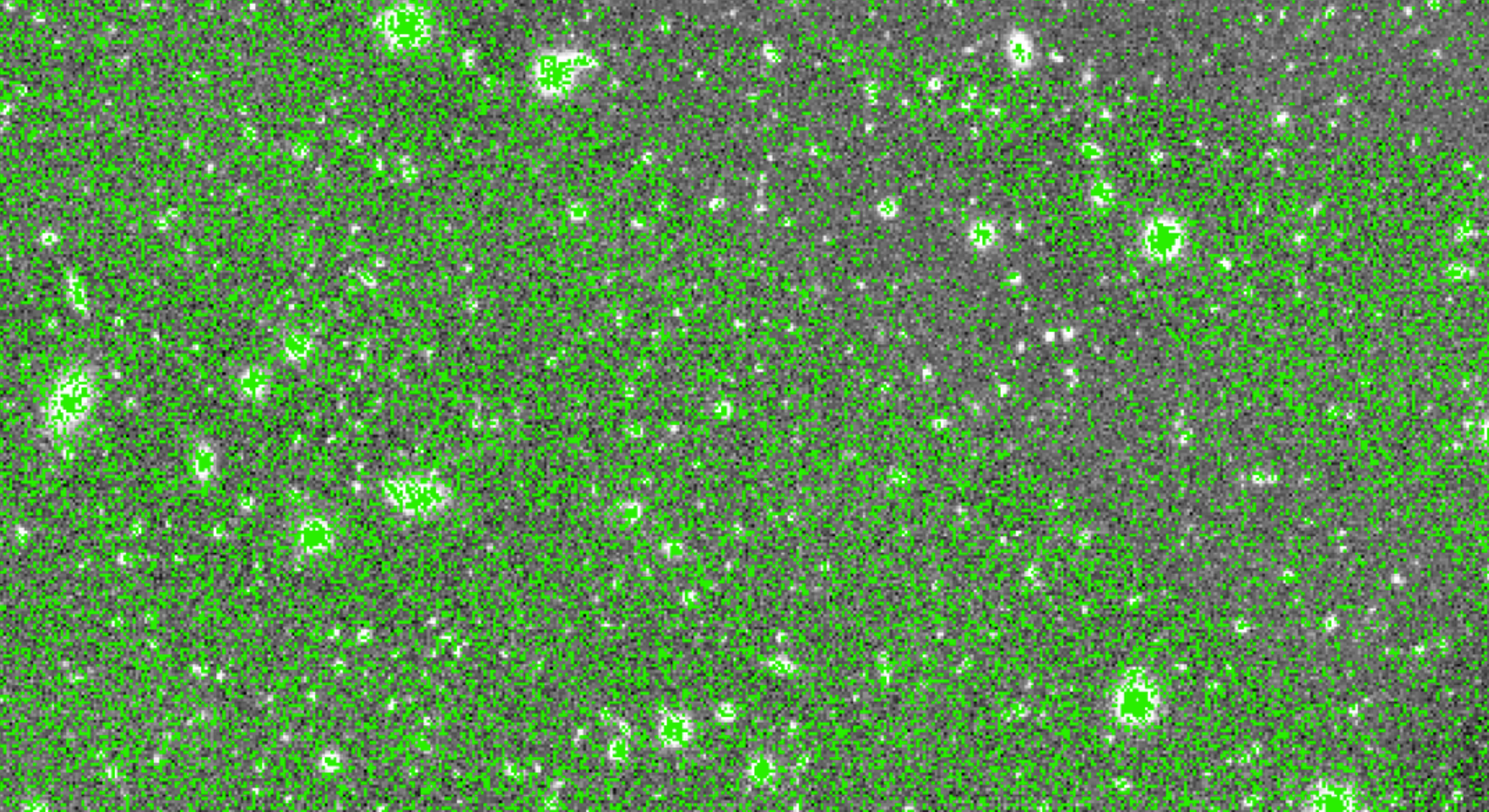}
 \label{fig:incr-3}
 }
\subfigure[]{
\includegraphics[width=0.20\linewidth]{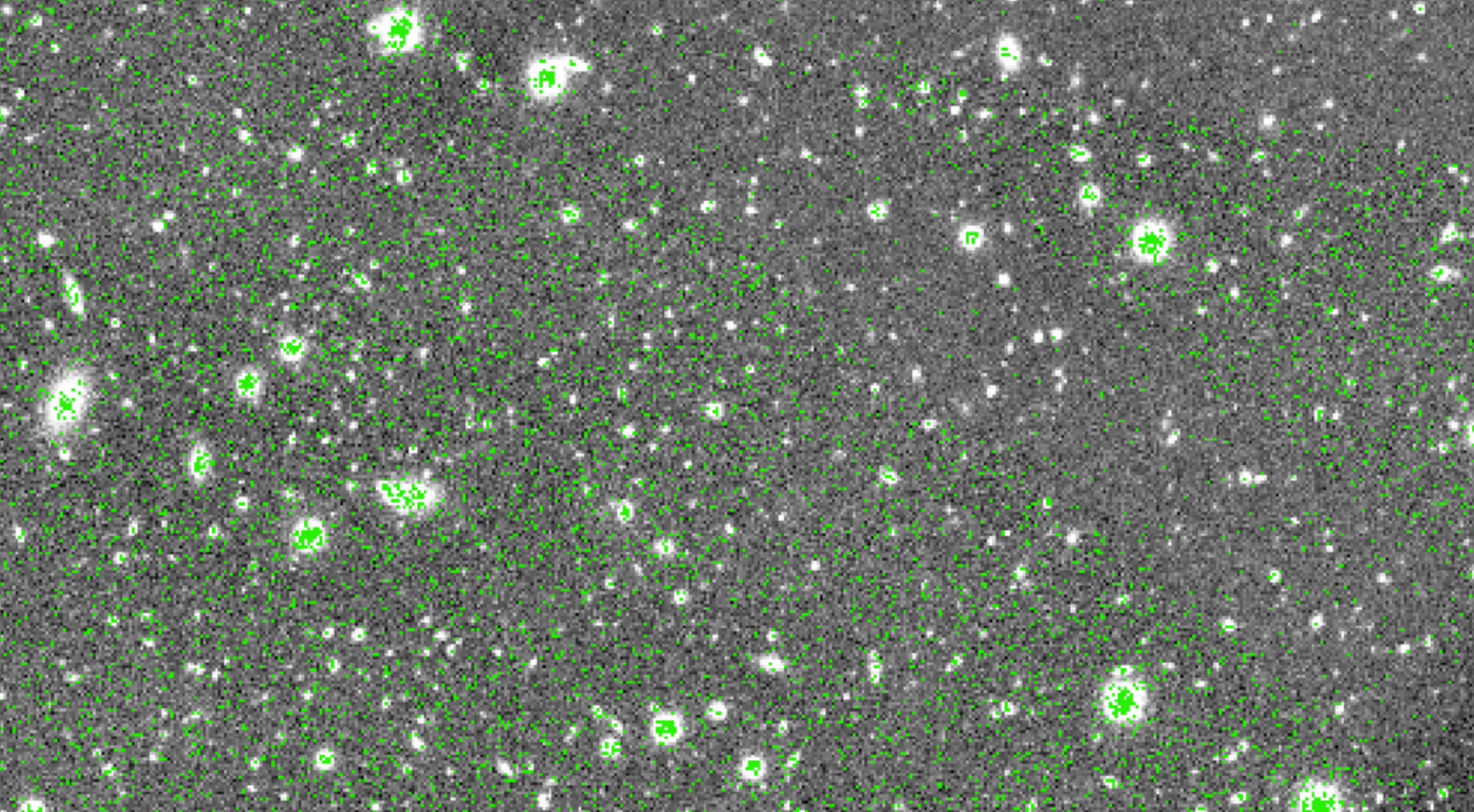}
 \label{fig:incr-4}
 }
%\subfigure[Iteration 4]{
%\includegraphics[width=0.4\linewidth]{img/incr-5.pdf}
% \label{fig:incr-5}
% }
%\subfigure[Iteration 5]{
%\includegraphics[width=0.4\linewidth]{img/incr-6.pdf}
% \label{fig:incr-6}
 %}
 \caption{Snapshots from the first 3 iterations of the \texttt{SigmaClip} application with the incremental-processing optimization on the LSST dataset. Green-colored points are the
 ones that change across iterations. (a) Original Image (b) Iteration-1 (c) Iteration-2 (d) Iteration-3.}
\label{fig:incr-lsst}
\end{center}
\vspace{-20pt}
\end{figure}

\begin{algorithm}[t] 
\tiny
\caption{\small ArrayLoop version of the \texttt{SigmaClip} application followed by image co-addition}
\label{alg:system-sigma-clip}
\begin{algorithmic}[1]
\renewcommand{\alglinenumber}[1]{\tiny #1.}
\Function{\textbf{ArrayLoop-sigma-clipping}}{$A$,$k$} \Comment{\texttt{SigmaClip} algorithm with FixPoint operator provided by the user.}
\State Input: Iterative Array $A$ $<$float $d$$>$[$x$,$y$,$t$], 
\State Input: $k$ a constant parameter.
\State $\pi: [x][y][z] \rightarrow [x][y].$
\State $\delta: ``A.d \ge \mu - k \times \sigma$ and $A.d \le \mu + k \times \sigma ? A : null "$ 
\State $f:\{avg() \;as\; \mu,stdv() \;as\; \sigma\}$
\State $FixPoint(A,\pi,f,\delta,count(),0)$ 
\EndFunction
\Statex
\Function{\textbf{ArrayLoop-incr-sigma-clipping}}{$A$,$k$} \Comment{ArrayLoop incremental rewriting of the \texttt{SigmaClip}.}
\State Input: Iterative Array $A$ $<$float $d$$>$[$x$,$y$,$t$], 
\State Input: $k$: a constant parameter.
\State Local: Iterative Array $C$ $<$int $c$,float $s$,float $s^2$$>$[$x$,$y$], 
\State Local: Array $S$ $<$float $\sigma$,float $\mu$$>$[$x$,$y$], 
\State $\Delta A^{-} \leftarrow A$ 
\While{($\Delta A^-$ is not empty)}
\State  \colorbox[gray]{0.90}{\parbox[t]{\dimexpr\linewidth-\algorithmicindent}{$T[x,y] \leftarrow$ select count$(d)$ as $c$,  sum($d$) as $s$, sum($d^2$) as $s^2$ from $\Delta A^-$ group by $x,y$\strut}} \label{alg:ext-dummy0}
\If{(first iteration)}
\State $C \leftarrow T[x,y]$ 
\Else 
\State  \colorbox[gray]{0.75}{merge($C$,$T$,$C.c - T.c$) \label{alg:ext-dummy1}} 
\State  \colorbox[gray]{0.75}{merge($C$,$T$,$C.s - T.s$)  \label{alg:ext-dummy2}}
\State  \colorbox[gray]{0.75}{merge($C$,$T$,$C.s^2 - T.s^2$)  \label{alg:ext-dummy3}}
\EndIf
\State  \colorbox[gray]{0.90}{\parbox[t]{\dimexpr\linewidth-\algorithmicindent}{$S[x,y] \leftarrow$ select $\frac{T.s}{T.c}$ AS $\mu$, $\sqrt[2]{\frac{T.s^2}{T.c} - (\frac{T.s}{T.c})^2}$ AS $\sigma$ FROM $\Delta^{+}C$\strut}}
\State  \colorbox[gray]{0.75}{merge$(A,S,S.\mu - k \times S.\sigma \le A.d \le S.\mu + k \times S.\sigma ? A : null)$}
%\State  $S[x,y,t]$ = SELECT $A.data$, $\Delta T^+.\mu$, $\Delta T^+.\sigma$ FROM $\Delta T^+$ join $A$ on $\Delta T^+.x$ = $A.x$ AND $\Delta T^+.y$=$A.y$
%\State $A[x,y,t]$ = SELECT $S.data$ FROM $S$ WHERE $S.data \ge S.\mu - k \times S.\sigma$ AND $S.data \le S.\mu + k \times S.\sigma$
\EndWhile
\EndFunction
%\Statex
\Statex \textbf{co-addition phase:}
\State $R[x,y] \leftarrow$ select sum($A.d$) as $coadd$ from $A$ group by $x,y$
\end{algorithmic}
\end{algorithm}

%ArrayLoop offers several optimization techniques to transform a na\"{\i}ve iterative query into an efficient query in order to speed-up the query runtime. 
%In this section we describe the basics of \textit{incremental iterative processing} in ArrayLoop and the extensions to the core SciDB. 
In a wide range of iterative algorithms, the output at each iteration
differs only partly from the output at the previous iteration.
Performance can thus significantly improve if the system computes, at
each iteration, only the part of the output that changes rather than
re-computing the entire result every time. This optimization called
\textit{incremental iterative processing}~\cite{ewen:12} is
well-known, \eg$\;$ in semi-naive datalog evaluation, and has been
shown to significantly improve performance in relational and graph
systems. ArrayLoop leverages the iterative computation model from
Section~\ref{sec:model} to automatically apply this optimization when
the semantics of the applications permit it.  The \texttt{SigmaClip}
application described in Section~\ref{ex:lsst} is an example
application that can benefit from incremental iterative processing.
Figure~\ref{fig:incr-lsst} shows multiple snapshots of running the
sigma-clipping algorithm using incremental iterative processing, on a
subset of the \texttt{lsst} dataset.  Green-colored points are the
ones with changed values across two consecutive iterations.  As the
iterative computation proceeds, the number of green-colored points
drops dramatically and consequently the amount of required computation
at that step.

\texttt{sigma-clipping()} and \texttt{incr-sigma-clipping()} modules
in Algorithm~\ref{alg:sigma-clip} show the original implementation and
the manually-written incremental version of the implementation,
respectively. In the \texttt{sigma-clipping()} module, the
\texttt{avg()} and \texttt{stdv()} aggregate operators are computed
over the whole input at each iteration, which is inefficient. In
\texttt{incr-sigma-clipping()}, the user rewrites the \texttt{avg()}
and \texttt{stdv()} aggregate operators in terms of two other
aggregate operators \texttt{count()} and \texttt{sum()}
(Algorithm~\ref{alg:sigma-clip}, Lines~\ref{line:incr1}
and~\ref{line:incr2}). The user also needs to carefully merge the
current partial aggregates with the aggregate result of the previous
iteration (Algorithm~\ref{alg:sigma-clip}, Line~\ref{line:merge1}). As
shown in Algorithm~\ref{alg:sigma-clip}, writing an efficient
incremental implementation is not a trivial task. It is painful for
users if they need to rewrite their algorithms to compute these
increments and manage them during the computation. Ideally, the user
wants to define the semantics of the algorithm and the system should
automatically generate an optimized, incremental implementation.
Additionally, as we show in the evaluation, if the system is aware of
the incremental processing, it can further optimize the implementation by
pushing certain optimizations all the way to the storage layer.

\subsection{Rewrite for Incremental Processing}

%figure out how to automatically rewrite input queries into an
%efficient incremental version of the algorithm.

%\magda{The following argument
%is not well phrased. It implies that a better algorithm exist but we
%  mean simply that pushing optimizations into the storage layer can
%  yield an even better runtime.} 
%In fact, the \texttt{incr-sigma-clipping()} module shown in
%algorithm~\ref{alg:sigma-clip} is not an optimized implementation. We
%can get better performance if we push some knowledge of increments all
%the way to the storage manager. 

In ArrayLoop, we show how the incremental processing optimization can
be applied to arrays. As shown in
Algorithm~\ref{alg:system-sigma-clip}, with ArrayLoop, the user
provides a \texttt{FixPoint} operator in
\texttt{ArrayLoop-sigma-clipping} function. ArrayLoop
\textit{automatically expands and rewrites the operation into an
  incremental implementation} as shown in the
\texttt{ArrayLoop-incr-sigma-clipping} function. The rewrite
  proceeds as follows. If the aggregate function $f$ is incremental,
  ArrayLoop replaces the initial aggregation with one over Delta A- or
  Delta A+ or both. For example, for \texttt{ArrayLoop-incr-sigma-clipping},
  only negative delta arrays are generated at each iteration(there is
  no $\Delta A^+$). So the rewrite produces a group-by aggregate only
  on $\Delta A^-$ (line 16). Next, ArrayLoop merges the partial
  aggregate values with the aggregate results from the previous
  iteration (lines 20 through 22). The aggregate rewrite rules define
  that merge logic for all the aggregate functions. In this example,
  ArrayLoop will generate one merge statement per aggregate function
  computed earlier. Finally, on Line 24, ArrayLoop does the final computation to generate the final aggregate
  values for this iteration. Note that finalize phase in the aggregate computation is always done on positive delta arrays ($\Delta C^+$), which generates the same result as computing on negative delta array $\Delta C^-$ followed by a subtract merge plus computing on positive delta array $\Delta C^+$ followed by an addition merge.  
  Line 25 leverages the $\delta$ function to generate the
  $\Delta A^-$ of the next iteration.

Currently the decision whether the application semantics permit incremental
iterative processing is left to the user.  Given the \texttt{FixPoint}
operator, ArrayLoop performs two tasks: (1) it automatically rewrites
aggregate functions, if possible, into incremental ones and (2) it
efficiently computes the last state of the iterative array using the
updated cells at each iteration. The automatic rewrite is enabled by
the precise model for iterative computations in the form of the three
functions $\pi$, $f$, and $\delta$. Given this precise specification
of the loop body, ArrayLoop rewrites the computation using a set of
rules that specify how to replace aggregates with their incremental
counter-parts when possible. To efficiently compute incremental state
updates, we introduce a special \textit{merge} operator. We now
describe both components of the approach.

%, described
%later. %, in the context of the array iterative model.

\textbf{(1) Automatic aggregate rewrite:} ArrayLoop triggers the
\textit{incremental iterative processing} optimization if any
aggregate function in the \texttt{FixPoint} operator is flagged as
incremental.  The Data cube paper~\cite{gray:97} defines an
aggregate function $F()$ as algebraic if there is an M-tuple valued
function $G()$ and a function $H()$ such that: $F(\{X_{i,j}\}) =
H(\{G(\{X_{i,j}\} |i=1,\dots, I\}) | j=1,\dots, J \})$.  ArrayLoop stores
a triple $(agg,\{G_1,\dots,G_k\},H)$ for any $algebraic$ function in
the system and rewrites the aggregate query in terms of $G()$ and
$H()$ functions during the query rewriting phase. For example, ArrayLoop
records the triple \texttt{(avg(),\{sum(),count()\},sum/count)} and
rewrites the \textit{algebraic} average function \texttt{avg()} using
the combination of \texttt{sum()} and \texttt{count()} to leverage
incremental iterative processing.

\textbf{(2) Incremental state management:} ArrayLoop provides an
efficient method for managing array state and incremental array
updates during the course of an iterative computation. We observe
that, during incremental processing, a common operation is to
\textit{merge} the data in two arrays, which do not necessarily have
the same number of dimensions. In our example application, merging
happens when the partial aggregates are combined with the aggregate
result of the previous iteration, line~\ref{line:merge1} in
\texttt{incr-sigma-clipping()} function. This operation merges
together two 2D arrays where the merge logic is inferred from the
incremental aggregate function $f$.  Such merging also happens when
the results of the aggregate function are used to update the iterative
array, lines~\ref{line:merge2} and~\ref{line:merge3} in
\texttt{incr-sigma-clipping()} function. In this case, the application
merges the data in a 2D array with the data in a 3D array by
\textit{sliding} or \textit{extruding} the 2D array through the 3D
array. The $\delta$ cell-update function defines the logic of the
merge operation in this case.  The $\pi$ assignment function pairs-up
cells from the intermediate aggregation array and the iterative array
that merge together and thus determines whether merging will
occur between arrays with the same number of dimensions or not.

In the manual implementation, shown in the
\texttt{incr-sigma-clipping()} function, the user implements the merge
logic manually using join and filter queries, which is inefficient.\footnote{From an engineering point of view,
  the new \textit{merge} operator, unlike a join, can also leverage
  vectorization where instead of merging one
  pair of matching cells at a time, ArrayLoop merges group of matching
  cells together, potentially improving query runtime, especially when
  the number of dimensions in the two input arrays is different.} To
remove this inefficiency, given the \texttt{FixPoint} operator, ArrayLoop
automatically generates queries with explicit merge points that
leverage a new merge operator that we add to SciDB:
\texttt{merge(Array source, Array extrusion, Expression exp)}.

\begin{figure}[t]
\begin{center}
% \subfigure[The source array $A_1$ and the extrusion array $A_2$.]{
%\includegraphics[width=0.2\linewidth]{img/extrusion-orig.pdf}
% \label{fig:ex-orig}
% }
 \subfigure[step 1]{
\includegraphics[width=0.15\linewidth]{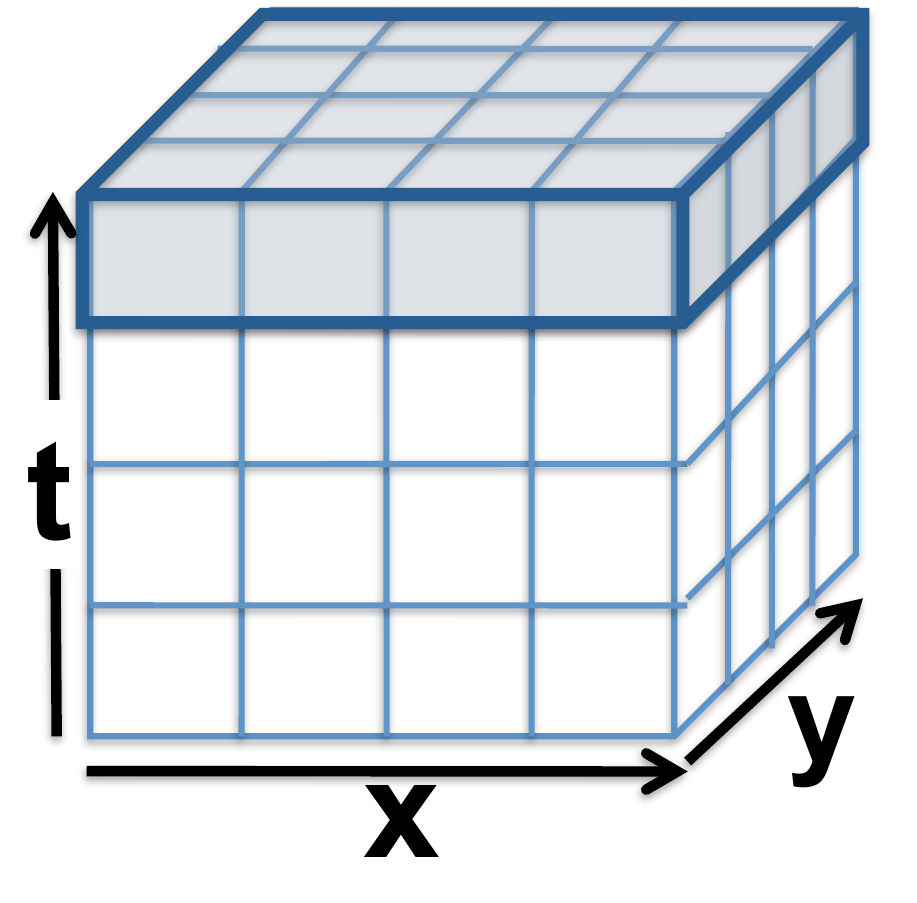}
 \label{fig:ex-1}
 }
\subfigure[step 2]{
\includegraphics[width=0.15\linewidth]{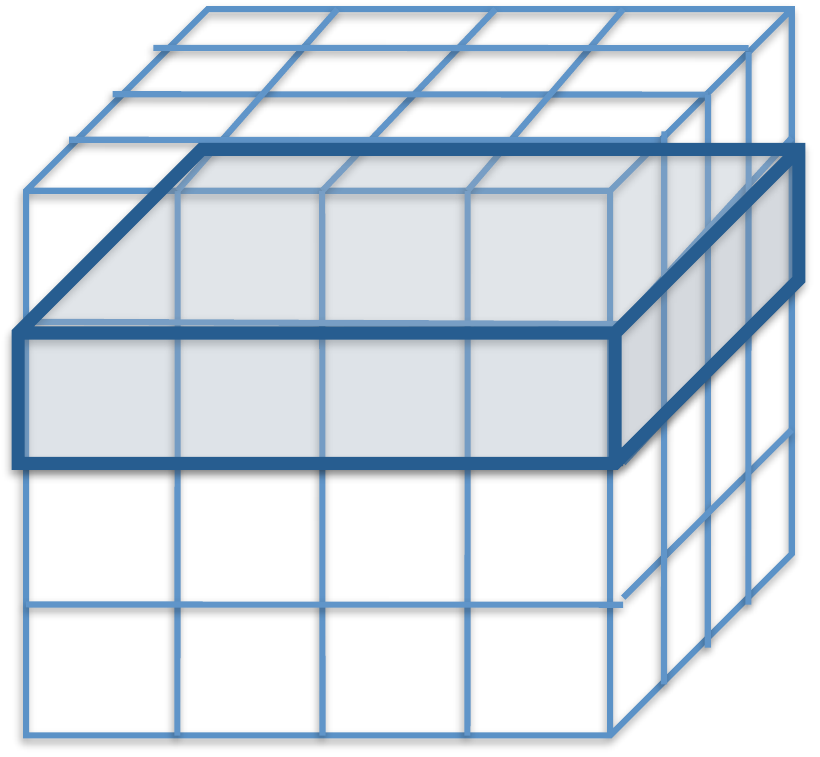}
\label{fig:ex-2}
 }
\subfigure[step 3]{
\includegraphics[width=0.15\linewidth]{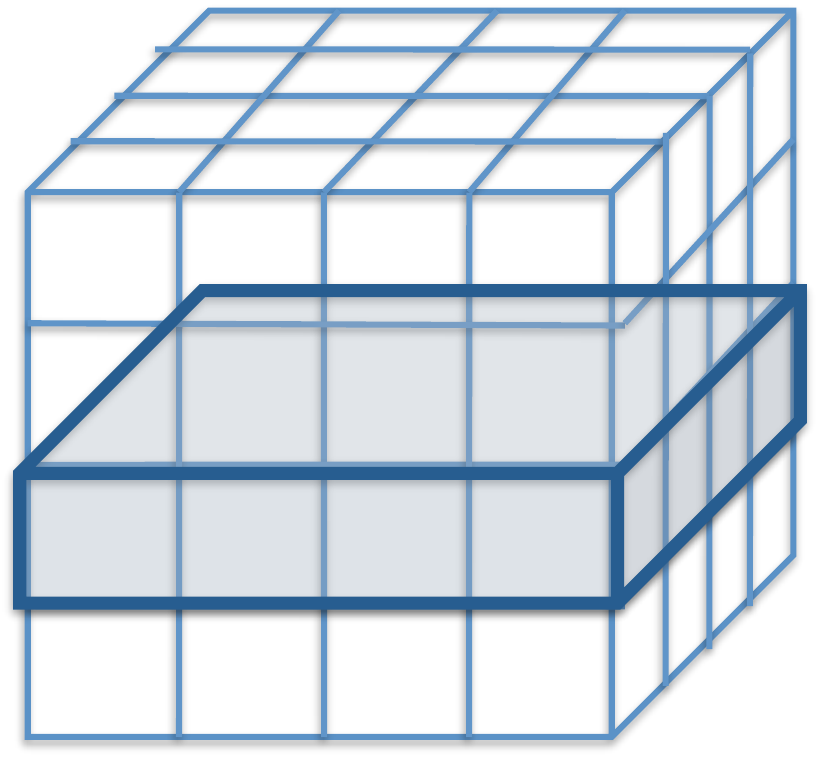}
 \label{fig:ex-3}
 }
 \subfigure[step 4]{
\includegraphics[width=0.15\linewidth]{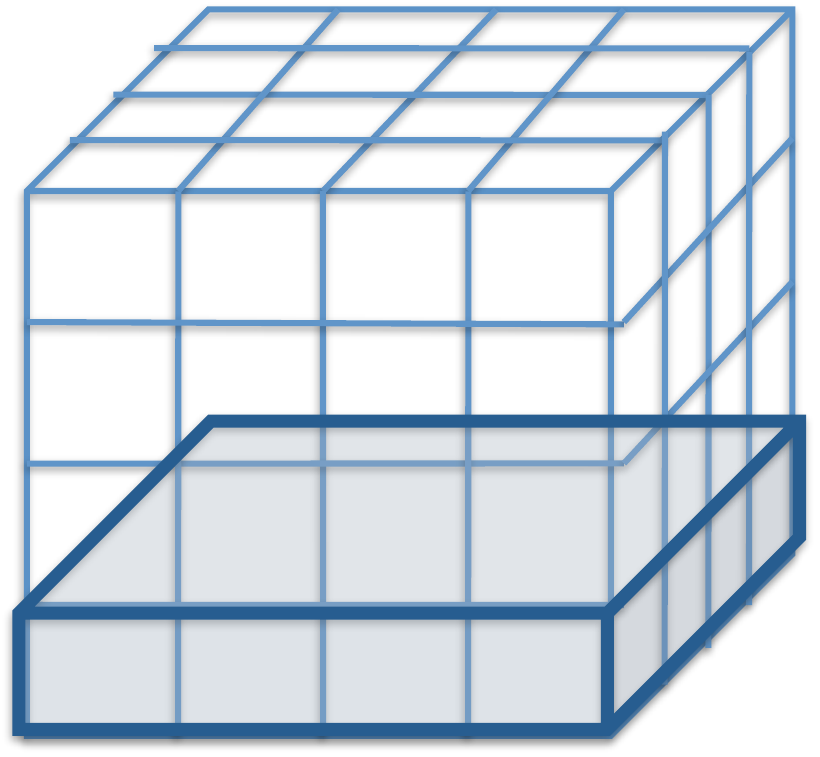}
 \label{fig:ex-4}
 }
 \caption[\texttt{merge()} operator example in \texttt{SigmaClip} application.]{merge$(A,T,T.\mu - k \times T.\sigma \le A.d \le T.\mu + k \times T.\sigma ? A : null)$ in \texttt{SigmaClip} application. This is the core filtering step where the outliers are removed. The 3D source array $A$ $<$float $d$$>$[$x$,$y$,$t$] and the 2D extrusion array (highlighted) $T$ $<$float $\mu$,float $\sigma$$>$[$x$,$y$] share the first two dimensions. (a), (b), (c), and (d) show how the cells in the extrusion array slide into the source array at runtime.}
\label{fig:extrusion}
\end{center}
\vspace{-20pt}
\end{figure}

The new \textit{merge} operator is unique in a sense that it
not only specifies the merge logic between two cells via a
mathematical expression, \texttt{exp}, but it also automatically
figures out which pairs of cells from the two arrays merge together by
examining their common dimensions. ArrayLoop merges two cells from the
\textit{source} array and \textit{extrusion} array if they share the
same dimension-values in those dimensions that match in
dimension-name. One cell in the \texttt{extrusion} array can thus
merge with many cells in the \texttt{source} array.
Figure~\ref{fig:extrusion} illustrates the merge operator for queries
in lines~\ref{line:merge2} and~\ref{line:merge3} in
Algorithm~\ref{alg:sigma-clip}. As the figure shows, the
\texttt{extrusion} array slides through the \texttt{source} array as
the merging proceeds.

\subsection{Pushing Incremental Computation into the Storage Manager}

We observe that increments between iterations translate
into updates to array cells and can thus be captured with two
auxiliary arrays: a \textit{positive delta array} and a
\textit{negative delta array}. At each iteration, the positive delta
array $\Delta A^+$ records the \textit{new} values of updated cells
and the negative delta array $\Delta A^-$ keeps track of the
\textit{old} values of updated cells. Delta arrays can
automatically be computed by the system directly at the
storage manager level.

As a further optimization, we extend the SciDB storage manager to
manage simple merge operations such as addition/subtraction found in
Lines~\ref{alg:ext-dummy1},~\ref{alg:ext-dummy2},
and~\ref{alg:ext-dummy3} of Algorithm~\ref{alg:system-sigma-clip}.  
ArrayLoop uses naming conventions as a
hint to the storage manager about the semantics of the merge
operation. For example $A_{(-)} \leftarrow B$, asks the storage
manager to subtract array $B$ from array $A$ and store the result of
the $(A-B)$ operation as the new version of array $A$.  In case array
$A$ is iterative, the new values and the old values of updated cells
are stored in $\Delta^{+} A$ and $\Delta^{-} A$, respectively.

Typically, the user need not worry about these annotations and
processing details since ArrayLoop automatically generates the
appropriate queries from the FixPoint specification.
However, the user can leverage these optimizations manually as
well. For example, the queries in the \texttt{incr-sigma-clipping()}
function at Lines~\ref{line:storage3},~\ref{line:storage4},
and~\ref{line:storage5} (queries with red box frames) can all be
pushed into the storage manager.

To achieve high performance, the storage manager keeps chunks of the
result array $A$ together on disk with the corresponding chunks from
the auxiliary $\Delta A^+$ and $\Delta A^-$ arrays.  As we showed in
previous work on array versioning~\cite{soroush:13a}, the space
overhead of delta arrays taken between similar array versions is
typically insignificant compared with the size of the original array.

We extend the \texttt{Scan()} and \texttt{Store()} operators to read
and write partial arrays $\Delta A^+$ and $\Delta A^-$,
respectively. With those optimizations, the user does not need to
explicitly write a user-defined \texttt{diff()} function or, as shown
in the \texttt{incr-sigma-clipping()} example, a sequence of
\texttt{join()} and \texttt{filter()} queries in order to extract
delta arrays from the output of the last iteration.

%\magda{For this figure, not clear what performance improvement is due
%  to the optimization shwon in this subsection vs the optimizations
%  shown in the subsection above.}  

In prior work~\cite{soroush:13b}, we demonstrated a prototype
implementation of the \texttt{SigmaClip} application together with the
incremental iterative processing optimizations in the context of the
AscotDB system that we built. AscotDB results from the integration of
ASCOT, a Web-based tool for the collaborative analysis of telescope
images and their metadata, and SciDB, a parallel array processing
engine. The focus of the demonstration was on this system integration
and on the impact of the optimizations on the application. AscotDB
shows that average users who use graphical interfaces to specify their
analysis also benefit from optimized iterative processing provided by
lower layers of the analysis stack.

\section{Iterative Overlap Processing}
\label{sec:overlap}

% \magda{Related work: We should cite Graphlab's ghost nodes in this
%   section. In the VLDB paper, they mention ghost nodes in
%   Sec. 4.1. They consider ghosts as caches and update them whenever
%   the value changes. Also, I thought that GraphLab included the
%   optimization where vertices are scheduled multiple times locally
%   before updating edges/vertices on other compute nodes but I do not
%   seem to find the passage describing this idea in any of theirimg
%   papers.}

% \emad{ TODO We should cite this: " ... Forexample, an update function
%   may choose to return (schedule) its neighbors only when it has
%   made a substantial change to its local data ... " this is similar
%   to mini-iteration. The difference is that update happens in graphlab
%   at the granularity of each node but in ours is per chunk ... also
%   each node can decide to schedule its neighbors independent of others
%   while in ours is a global decision ( based on some aggregated value
%   of changes in the whole array. }

To process a query over a large-scale array in parallel, SciDB (and
other engines) break arrays into sub-arrays called chunks, distribute
chunks to different compute nodes (each node receives multiple
chunks), and process chunks in parallel at these nodes. For many
operations, such as filter for example, one can process chunks
independently of each other and can union the result.  This simple
strategy, however, does not work for many scientific array
operations. Frequently, the value of each output array cell is based
on a neighborhood of input array cells. Data clustering is one
example. Clusters can be arbitrarily large and can go across array
chunk boundaries. A common approach to computing such operations in
parallel is to perform them in two steps: a local, parallel step
followed by an aggregate-type post-processing
step~\cite{kwon:10a,kwon:10b,mahout} that merges partial results into
a final output. For the clustering example, the first step finds
clusters in each chunk. The second step combines clusters that cross
chunk boundaries~\cite{kwon:10a}.  Such a post-processing phase,
however, can add significant overhead. To avoid a post-processing
phase, some have suggested to extract, for each array chunk, an
overlap area $\epsilon$ from neighboring chunks, store the overlap
together with the original chunk~\cite{rogers:10,seamons:94}, and
provide both the core data and overlap data to the operator during
processing~\cite{soroush:11a}. This technique is called
\textit{overlap processing}. Figure~\ref{fig:mini-iteration} shows an
example of array chunks with overlap. We refer the interested reader
to our ArrayStore paper~\cite{soroush:11a} for a more detailed
discussion of efficient overlap processing techniques. These
techniques, however, do not address the question of how best to update
the overlap data during an iterative computation. Our contribution in
this paper is to tackle this specific question.

%In
%the ArrayStore paper~\cite{soroush:11a}, the authors discussed
%different strategies to support efficient overlap processing.

% In a parallel array processing engine, the ideal is to process each
% array fragments (or chunks) independent of the others and simply union
% the results.  However, this simple strategy is not applicable to many
% scientific array operations. In fact, operations such as clustering
% require that an operator considers data from a range of neighboring
% cells in order to produce each output cell. These neighborhoods are
% often called ``stencils" and they are bounded in size. A common
% approach to run these problems in parallel is to perform a
% post-processing step~\cite{mahout,kwon:10b,kwon:10a} that merges
% partial results in each fragment and returns the final
% output. However, such a post-processing phase can add significant
% overhead. To avoid a post-processing phase, some have suggested to
% extract, for each array chunk, an overlap area $\epsilon$ from
% neighboring chunks, store the overlap together with the original
% chunk~\cite{rogers:10,seamons:94}, and provide both to the operator
% during processing. This technique is called \textit{overlap
%   processing}. In the ArrayStore paper~\cite{soroush:11a}, the authors
% discussed different strategies to support efficient overlap
% processing.

\subsection{Efficient Overlap Processing}

%Overlap processing can be especially helpful for iterative
%computations that need to perform an operation multiple times until a
%termination condition.  

%A technique similar to \textit{overlap
 % processing} has been used in iterative graph-processing
%systems. 

% This should move to the related work
%Distributed GraphLab~\cite{low:12} proposed \textit{ghost
 % nodes} that are replicas of vertices across graph partitions. The
%idea is to provide each vertex in the graph with direct memory access
%to all its neighboring vertices. The challenge is then to efficiently
%keep the cached vertices up to date.  Similarly, ArrayLoop leverages
%overlapping techniques to support iterative parallel array processing,%
%where cells at boundary of chunk partitions are replicated and must be
%kept up to date.

Array applications that can benefit from overlap processing techniques
are those that update the value of certain array cells by using the
values of neighboring array cells. 
%In terms of our proposed array
%iterative model described in Section~\ref{sec:model}, those are
%applications with one-to-one \textit{window} assignment functions $\pi$. 
The \texttt{SourceDetect} application described in
Section~\ref{ex:detection} is an example application that can benefit
form overlap processing. Other example applications include
``oceanography particle tracking'', which follow a set of particles as
they move in a 2D or 3D grid. A velocity vector is associated with
each cell in the grid and the goal is to find a set of trajectories,
one for each particle in the array. Particles cannot move more than a
certain maximum distance (depending on the maximum velocity of
particles) at each step.  These applications can be effectively
processed in parallel by leveraging overlap processing techniques.

The challenge, however, is to keep replicated overlap cells up-to-date
as their values change across iterations. To efficiently update
overlap array cells, we leverage SciDB's bulk data-shuffling operators
as follows: SciDB's operator framework implements a \texttt{bool
  requiresRepart()} function that helps the optimizer to decide
whether the input array requires repartitioning before the operator
actually executes. The partitioning strategy is determined by the
operator semantics. For example, \texttt{WindowAggregate}
operator~\cite{scidbguide} in SciDB requires repartitioning with
overlap in case the input array is not already partitioned in that
manner. We extend the SciDB operator interface such that ArrayLoop can
dynamically set the returned value of the operator's
\texttt{requiresRepart()} function. To update overlap data, ArrayLoop
sets the \texttt{requiresRepart()} return value to true. ArrayLoop has
the flexiblity to set the value to true either at each iteration or
every few iterations as we discuss further below. In
case an operator in SciDB is guided by ArrayLoop to request
repartitioning, the SciDB optimizer injects the
\texttt{Scatter/Gather}~\cite{scidbguide} operators to shuffle the
data in the input iterative array before the operator
executes. 

%With this
%approach, ArrayLoop leverages the rich set of array operators
%available in SciDB to keep the overlap data up to date.  One benefit

%of this approach is that array operators in SciDB runs at the chunk
%level.  They receive data chunks as inputs and produce a data chunk as
%output. Therefore, the \texttt{Scatter/Gather} operation re-shuffles

A benefit of using SciDB's existing \texttt{Scatter/Gather} operators,
is that they shuffle array data one chunk (\ie, sub-array) at a
time. Chunk-based data shuffling is faster compared with the method
that shuffles overlap data one cell at a time. The downside of using
SciDB's \texttt{Scatter/Gather} general operators is the relative
higher cost of data shuffling when only a few overlap cells have
changed.

\subsection{Mini-Iteration Processing}

Keeping overlap cells updated at each iteration requires
 reading data from disk, shuffling it across the network,
and writing it to disk.  These are all expensive operations.  Any
reduction in the number of such data synchronization steps can yield
significant performance improvements.

We observe that a large subset of iterative applications have the
property that overlap cells can be updated only every few
iterations. These are applications, for example, that try to find
structures in the array data, \eg~\texttt{SourceDetect}
application. These applications can find structures locally and
eventually need to exchange information to stitch these local
structures into larger ones. For those applications, ArrayLoop
can add the
following additional optimization: ArrayLoop runs the algorithm for multiple
iterations without updating the replicas of overlap cells. The
application iterates over chunks locally and independently of other
chunks. Every few iterations, ArrayLoop triggers the update
of overlap cells, and continues
with another set of local iterations.  The key idea behind this
approach is to avoid data movement across array chunks unless 
a large enough amount of change justifies the cost.

\begin{figure}[t]
\centering
\includegraphics[width=0.60\linewidth]{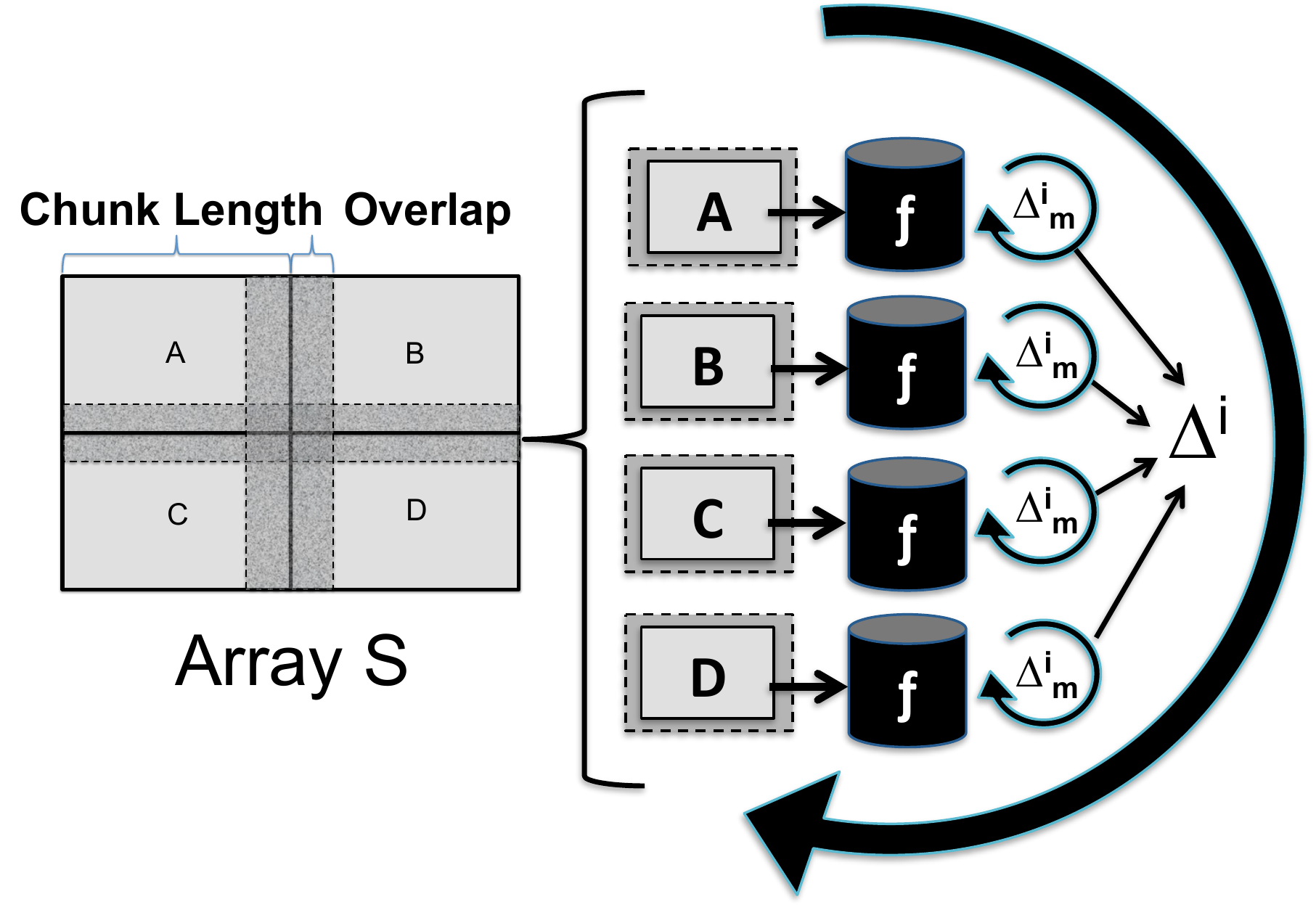}
\caption{Illustration of the mini-iteration optimization. $\Delta^{i}_{m}$ represents local changes 
at iteration $i$ of mini-iteration $m$ and $\Delta^i$ is the global change at iteration $i$. }
\label{fig:mini-iteration}
\vspace{-10pt}
\end{figure}

We call each series of local iterations without overlap cell
synchronization \textit{mini
  iterations}. Figure~\ref{fig:mini-iteration} illustrates the
schematic of the \text{mini iteration optimization}. 

% This is not exactly similar. We talk about synchronizing
% overlap data not executing updates. In any case, this 
% should move to related work.
%A similar idea
%has already been exercised in other data management systems with
%i%teration support. For example, in distributed GraphLab~\cite{low:12},
%a% vertex may choose to schedule its neighbors only when it has made a
%s%ubstantial change to its local data. We borrow this optimization and
%a%pply it to arrays to measure the efficiency of that optimization in
%this new setting. Unlike GraphLab, ArrayLoop makes the overlap data
%shuffling a global decision for the entire array not the individual cells,
%which leverages SciDB's simpler scheduler and also amortizes the cost
%by synchronizing all overlap cells in the same operation. 

An alternative approach is for the scheduler to delegate the decision
to shuffle overlap data to individual chunks, rather than making the
decision array-global as we do in this paper. We leave this extra
optimization for future work.

%or cell-local as in
%GraphLab. 

% In algorithms such as the one in \texttt{SourceDetect} application
% where the overlapped cells synchronization is needed, one possible
% optimization is to run the iterative computation locally at each chunk
% until \textit{local convergence} occurs on all the chunks and then
% trigger the expensive overlap data shuffling. 

% The idea behind mini-iterations is to avoid data movement across nodes
% unless we are sure there are enough changes to justify the cost. We
% call each of those iterations that run locally at each chunk with no
% overlapping cells synchronization barrier at the end a \textit{mini
%   iteration}. Figure~\ref{fig:mini-iteration} illustrates schematic of
% the \text{mini iteration optimization}. A similar idea has been
% exercised in other data management systems with iteration support as
% well. For example, in distributed GraphLab~\cite{low:12}, a vertex may
% choose to schedule its neighbors only when it has made a substantial
% change to its local data. We borrow this optimization and apply it to
% arrays to measure the efficiency of that optimization in this new
% setting. Unlike Graphlab, ArrayLoop made the overlap data shuffling a
% global decision for the entire cells in the array not each single
% cell, which result in a simpler scheduler in SciDB. An alternative
% approach is to choose an in-between solution where the scheduler
% delegates the decision to shuffle overlap data to each chunk, nor as
% coarse-grained as the entire array, neither as fine-grained as a
% single overlapped cell. We do not explore this optimization in this
% paper.

\begin{figure}[t]
\centering
\includegraphics[width=0.40\linewidth]{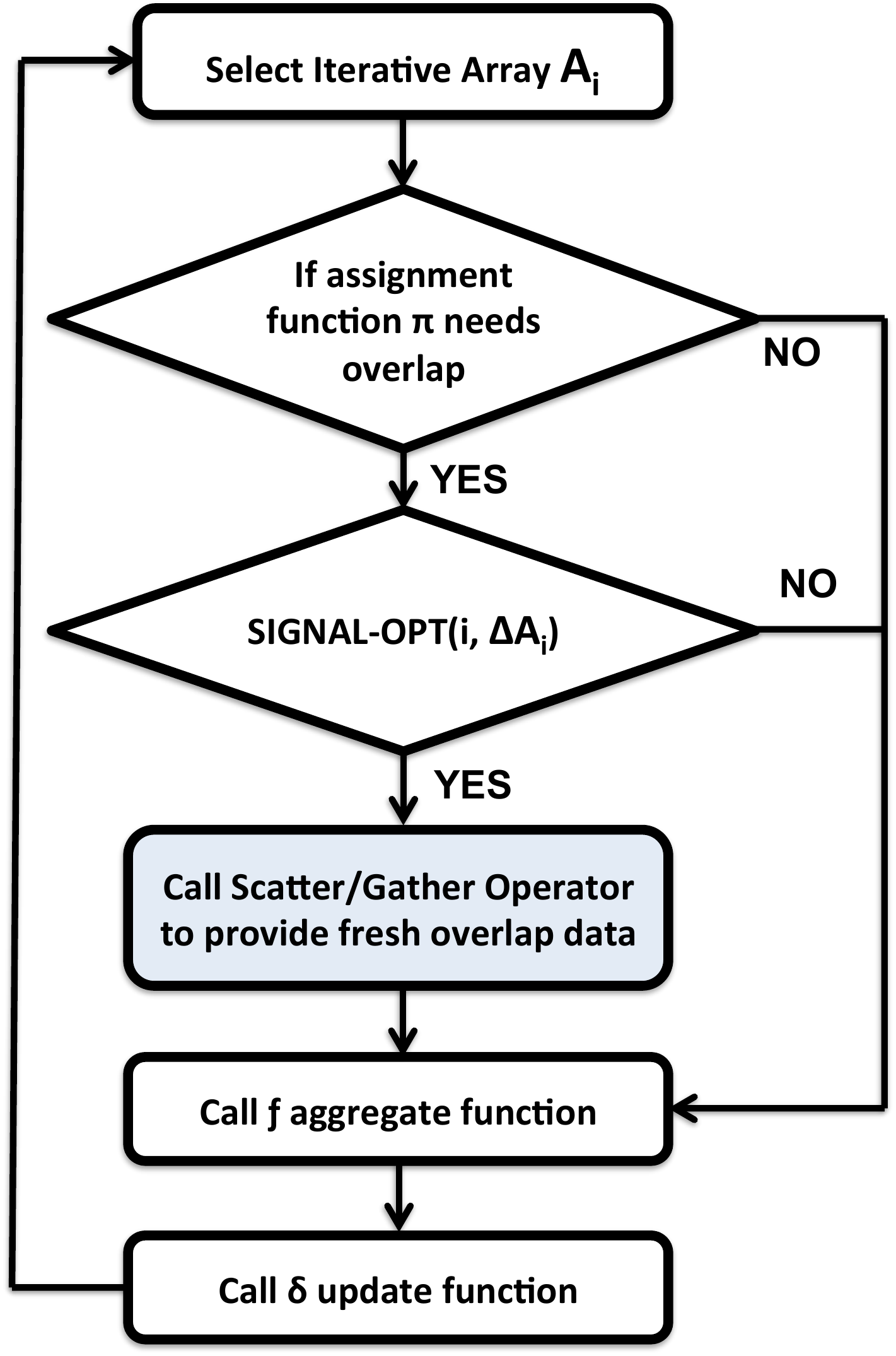}
\caption{Control flow diagram for mini-iteration-based processing in ArrayLoop.}
\label{fig:flow}
\vspace{-15pt}
\end{figure}

ArrayLoop includes a system-configurable function \texttt{SIGNAL-OPT()}
that takes as input an iteration number and a delta iterative array,
which represents the changes in the last iteration. This function is
called at the beginning of each iteration. The output of this function
defines if the overlap data at the current iteration needs to be
shuffled. A control flow diagram of this procedure is shown in
Figure~\ref{fig:flow}.  There exists an optimization opportunity to
exploit: Do we exchange overlap cells every iteration? Or do we wait
until local convergence? Or something in between these two extremes?
We further examine those optimization questions in
Section~\ref{sec:eval}.

%\begin{algorithm}[H]
%\caption{\texttt{Sync-Overlap} function to synchronize overlap cells at the end of each iteration.}
%\label{alg:syn-overlap}
%\begin{algorithmic}[1]
%\Function{Sync-Overlap}{$A$,$\Delta A^+$}
%\State $A$: Iterative array.
%\State $\Delta A^+$: positive delta array $A$ at the current iteration.
%\State $\Delta A_{flat}$ = unpack($\Delta A^+$)
%\State $\Delta A_{m}$ = redimension($\Delta A_{flat}$,$A.schema$)
%\State merge($A$,$\Delta A_{m}$,($\Delta A_{m}$.d$\ne \phi$):$\Delta A_{m}$.d$?A$.d)
%\EndFunction
%\end{algorithmic}
%\end{algorithm}
%
%\begin{algorithm}[H]
%\caption{Mini-iteration optimization}
%\label{alg:mini-iteration}
%\begin{algorithmic}[1]
%\Statex \textbf{input:} $A$: Iterative array, $f:$ local iterative function, $\delta:$ update function, $\Delta:$ delta collector array. 
%\While{$T(\Delta) \le \epsilon$}
%\State $m=0$
%\State $\Delta \leftarrow \phi$ \Comment{delta collector}
%\While{$T(A_{i,m}) \le \epsilon$} \Comment local convergence
%\State $A_{i,m} = \delta(A_{i,m},f(A_{i,m}))$
%\State merge($\Delta$,$\Delta A^{+}_{i,m}$,($\Delta A^{+}_{i,m}$.d$\ne \phi$):$\Delta A^{+}_{i,m}$.d$?D$.d)
%\State m++
%\EndWhile
%\State Sync-Overlap($A$,$\Delta$)
%\State i++
%\EndWhile
%\end{algorithmic}
%\end{algorithm}

\section{Multi-Resolution Optimization}
\label{sec:multi-res}

In many scientific applications, raw data lives in a continuous space
(3D universe, 2D ocean, N-D space of continuous variables).
Scientists often perform continuous measurements over the raw data and then
store a discretized approximation of the real data in arrays.  In
these scenarios, different levels of granularity for arrays are
possible and scientifically meaningful to analyze. In fact, it is
common for scientists to look at the data at different levels of
detail.

As discussed earlier, many algorithms search for structure
in array data. One example is the extraction of celestial objects from
telescope images, snow cover regions from satellite images, or
clusters from an N-D dataset. In these algorithms, it is often
efficient to first identify the outlines of the structures on a
low-resolution array, and then refine the details on high-resolution
arrays. We call this array-specific optimization
\textit{multi-resolution} optimization. This multi-resolution
optimization is a form of prioritized processing. By first processing
a low-resolution approximation of the data, we focus on identifying
and approximating the overall shape of the structures. Further
processing of higher-resolution arrays helps extract the more
detailed outlines of these structures.

In the rest of this section we describe how ArrayLoop automates this
optimization for \textit{iterative} computations in SciDB. We use the
\texttt{KMeans} application described in Section~\ref{ex:kmeans} and
the \texttt{SourceDetect} application described in
Section~\ref{ex:detection} as our illustrative examples.

\begin{figure}[t]
\centering
\includegraphics[width=0.6\linewidth]{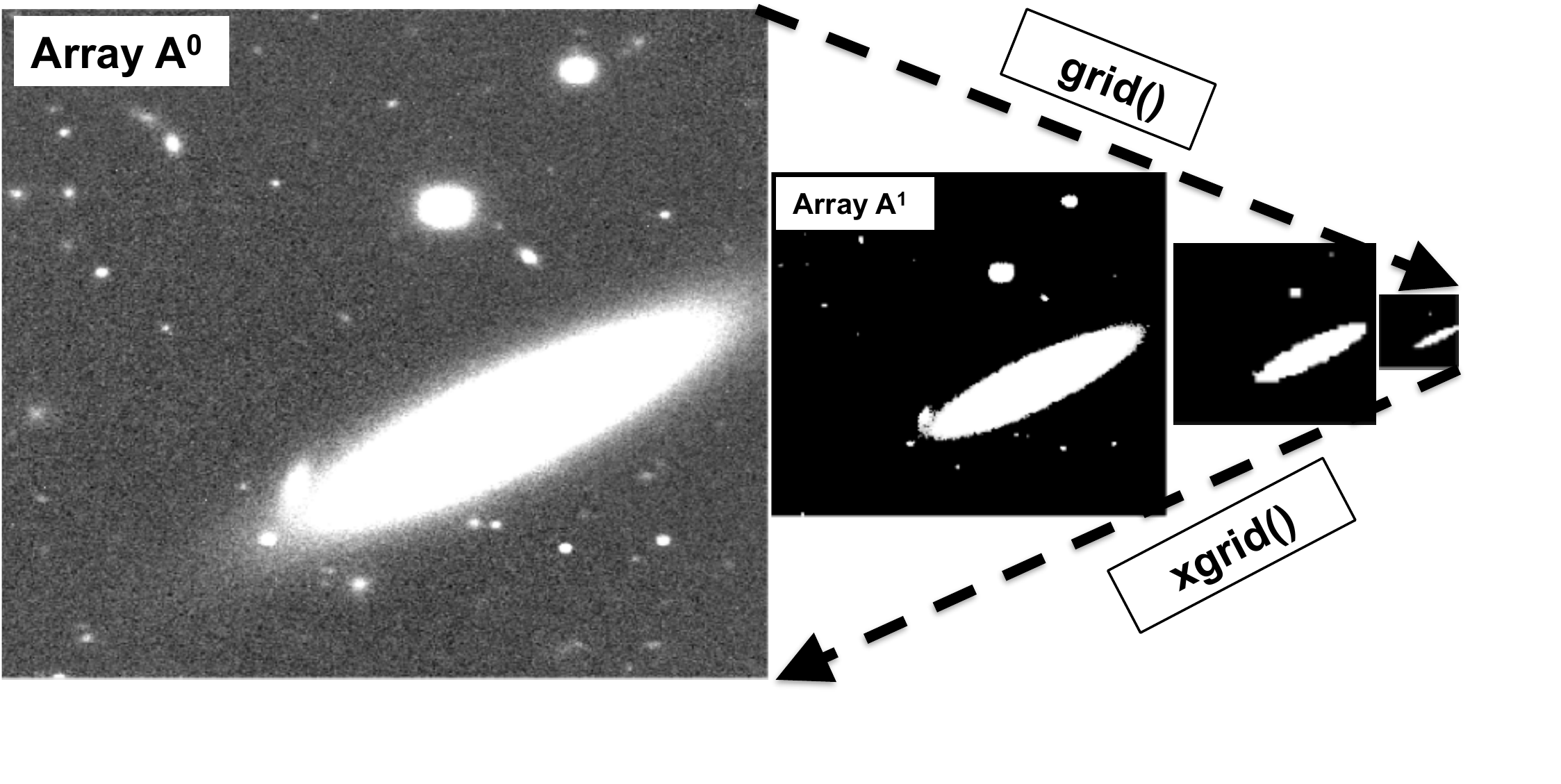}
\vspace{-0.5cm}
\caption[Illustration of the multi-resolution optimization for the \texttt{SourceDetect} application.]{Illustration of the multi-resolution optimization for the \texttt{SourceDetect} application. There is a sequence of three grid operations initiated from the original \texttt{lsst} image $A^0$: $A^{0} \xrightarrow{grid_p(A^{0},2,2)} A^{1} \xrightarrow{grid_p(A^{1},2,2)} A^{2} \xrightarrow{grid_p(A^{2},2,2)}  A^3$.}
\label{fig:multi-res-schematic}
\vspace{-10pt}
\end{figure}

To initiate the \textit{multi-resolution} optimization, ArrayLoop
initially generates a series of versions, $A^{i}, A^{i+1}, \dots ,
A^{j}$, of the original iterative array $A$. Each version has a
different resolution. $A^{i}$ is the original array. It has the
highest resolution. $A^{j}$ is the lowest-resolution
array. Figure~\ref{fig:multi-res-schematic} illustrates three
pixelated versions of an \texttt{lsst} image represented as iterative
array $A^0$ in the context of the \texttt{SourceDetect}
application. The coarser-grained, pixelated versions are generated by
applying a sequence of \texttt{grid} followed by \texttt{filter}
operations represented together as \texttt{$grid_{p}$()}, where $p$ is
the predicate of the \texttt{filter} operator.  The size and the
aggregate function in the \texttt{grid} operator are
application-specific and are specified by the
user. The \texttt{SourceDetect} application has a grid-size of
$(2\times2)$ and an aggregate function \texttt{count} with a filter
predicate that only passes grid blocks without empty cells (in this
scenario all the grid blocks with count=4).  This ensures that cells
that are identified to be in the same cluster in a coarsened version
of the array, remain together in finer grained versions of the array
as well.  In other words, the output of the iterative algorithm on the
pixelated version array $A^{j}$ should be a \textit{valid}
intermediate step for $A^{j-1}$.  ArrayLoop runs the iterative function
$Q$ on the sequence of pixelated arrays in order
of increasing resolution. The output of the iterative algorithm after
convergence at pixelated version $A^{i}$ is transformed into a
finer-resolution version using an \texttt{xgrid} operator (inverse of
a grid operator). It is then merged with $A^{i-1}$, the next immediate
finer-grained version of the iterative array. We represent both
operations as \texttt{$xgrid_{m}$()}. The \texttt{xgrid}
operator~\cite{scidbguide} produces a result array by scaling up its
input array. Within each dimension, the \texttt{xgrid} operator
duplicates each cell a specified number of times before moving to the
next cell. The following illustrates the ordered list of
operators called by ArrayLoop during \textit{multi-resolution}
optimizations:
\par\nobreak
{\scriptsize  
  \setlength{\abovedisplayskip}{6pt}
  \setlength{\belowdisplayskip}{\abovedisplayskip}
  \setlength{\abovedisplayshortskip}{0pt}
  \setlength{\belowdisplayshortskip}{3pt}
\begin{align}
\label{eq:multi-res}
\tiny 
&A^{0} \xrightarrow{grid_p()} \dots A^{i} \xrightarrow{grid_p()} A^{i+1} \xrightarrow{grid_p()} \dots A^j \nonumber \\
&A^j \xrightarrow{Q} A^{*j} \xrightarrow{xgrid_m(A^{*j})} A_{x}^{j-1} \nonumber \\
&\dots \\
&A_{x}^{1} \xrightarrow{Q} A^{*1} \xrightarrow{xgrid_m(A^{*1})} A_{x}^{0} \nonumber \\
&A_{x}^{0} \xrightarrow{Q} A^{*0} \nonumber
\end{align} 
\vspace{-10pt}
}%
%Now we clarify the idea behind this optimization by a simple iterative
%source detection example \magda{When we introduced the example, we did
%  not introduce $\epsilon$.}. One way to implement this use-case is
%described at Algorithm~\ref{alg:sd-sq}. Initially all the points in
%the original array are uniquely labeled. The body of the loop is as
%follow: scan the entire array with a ($2\epsilon \times 2\epsilon$)
%window and at each step during the scan take the minimum of all the
%integer labels inside the window and update the cell at the center of
%the window with the output of the
%aggregation. 

where $A^{*i}$ is the output of the iterative algorithm $Q$ on
pixelated array $A^{i}$, and $A^{j-1}$ is replaced with $A_{x}^{j-1}$
as the new input for the iterative computation at pixelated version $(j-1)$.

By carefully merging the approximate
results with the input array at the next finer-grained level, ArrayLoop
skips a significant amount of computation.

%By applying the multi-resolution optimization,
%ArrayLoop initially aims to find course-grained outlines of the clusters
%in more pixelated versions and delays the more detailed findings for
%finer-grained versions. 

The K-means clustering algorithm on points in a continuous space is another example application that
benefits from this optimization. The \texttt{KMeans} application can
use an arbitrary grid size. It also uses \texttt{count} as the
aggregate function with a filter predicate that passes grid blocks
that have at least one non-empty cell. It is easy to observe 
that in case of K-means clustering, $A_{x}^{j-1}$ is a \textit{valid}
labeling for the next pixelated array $A^{j-1}$. Basically, K-means clustering on
$A^{j}$ produces a set of centroids for the k-means algorithm on
$A^{j-1}$ that lead to a faster convergence than a random set of initial centroids.

The advantage of applying the \textit{multi-resolution} optimization
goes beyond better query runtime performance. This optimization can
also help when the original iterative array changes, which is described
as the following additional optimization:

\paragraph*{Input Change Optimization} If ArrayLoop materializes the
outputs $A^{*i}$ for all the pixelated versions of the original array
$A$, then there is an interesting optimization in case the
 original iterative array $A$ is modified. Unlike the Naiad system~\cite{mcsherry:13} that
materializes the entire state at each iteration to skip some
computation in case of change in the input data, ArrayLoop takes a
different strategy. When changes in the input occur, ArrayLoop  re-generates
the pixelated arrays $A^{i}$s in Equation~\ref{eq:multi-res}, but only
runs the iterative algorithm $Q$ for those arrays $A^{i}$s that have
also changed in response to the input array change.
If $A^{i}$ did not change
for some $i$, ArrayLoop skips the computation $A^k \xrightarrow{Q} A^{*k}
\; \forall k \ge i$ and uses the materialized result $A^{*i}$ from the
previous run to produce $A_{x}^{i-1}$. The intuition is that, if there
are only a few changes in the input array, it is likely that
changes are not carried over to all the pixelated versions of the
array and our system reuses some results of the previous run for
the current computation as well.

\section{Evaluation}
\label{sec:eval}

In this section, we demonstrate the effectiveness of ArrayLoop's
native iterative processing capabilities including the three
optimizations on experiments with 1TB of \texttt{LSST}
images~\cite{repository}. Because the LSST will only start to produce
data in 2019, astronomers are testing their analysis pipelines with
synthetic images that simulate as realistically as possible what the
survey will produce. We use one such synthetic dataset. The images
take the form of one large 3D array (2D images accumulated over time)
with almost 44 billion non-empty cells. The experiments are executed
on a 20-machine cluster.  (Intel(R) Xeon(R) CPU E5-2430L @ 2.00GHz)
with 64GB of memory and Ubuntu 13.04 as the operating system. We
report performance for two real-scientific applications
\texttt{SigmaClip} and \texttt{SourceDetect} described in
Sections~\ref{ex:lsst} and~\ref{ex:detection},
respectively. \texttt{SigmaClip} runs on the large 3D array and
\texttt{SourceDetect} runs on the co-added 2D version of the whole
dataset.

\subsection{Incremental Iterative Processing}

\begin{figure}[t]
\centering
  \subfigure[\texttt{SigmaClip} application with different
  strategies.]
  {\includegraphics[width=0.80\linewidth]{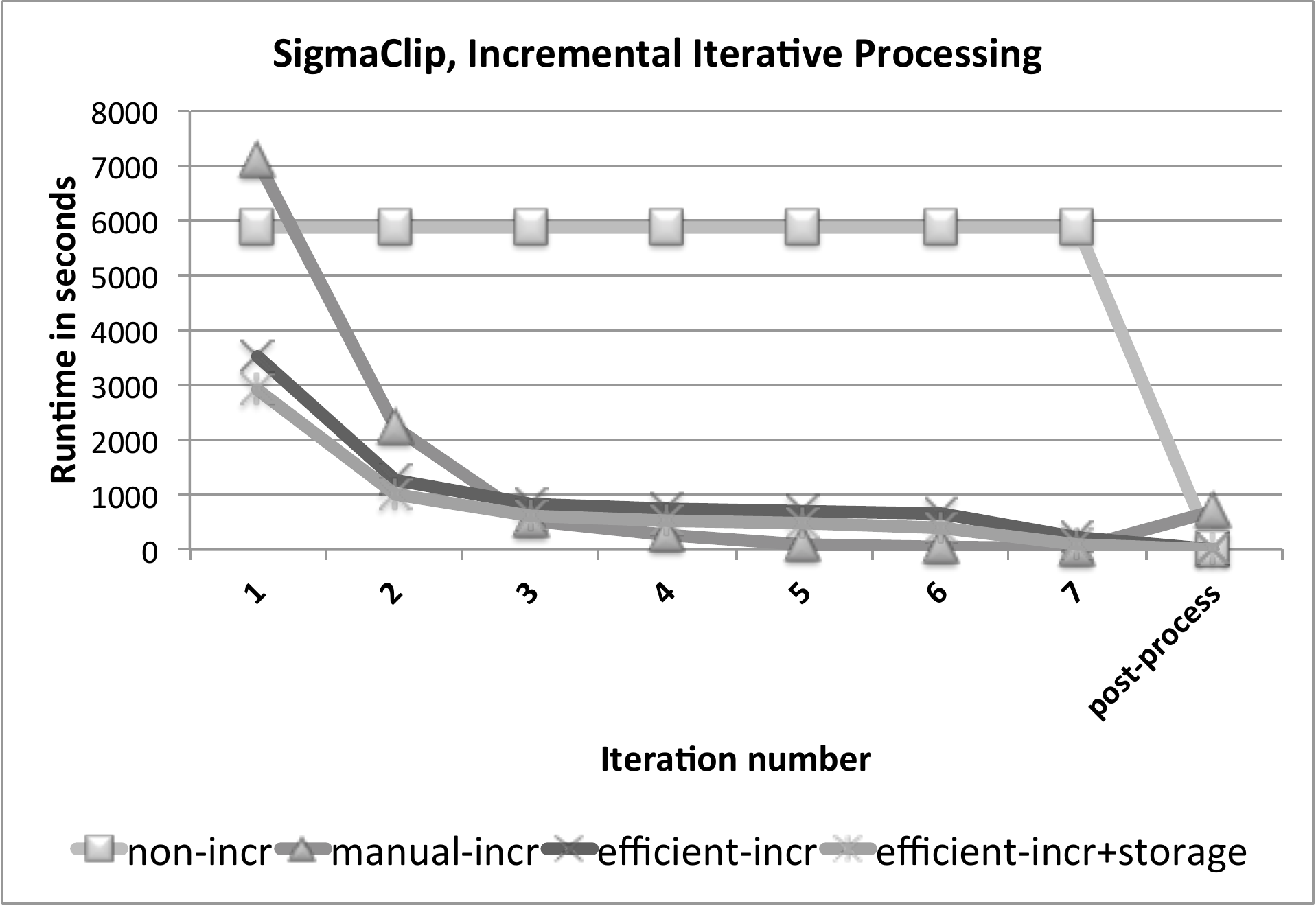}
\label{fig:SigmaClip-incr}
}
\subtable[Total runtime for \texttt{SigmaClip} for different strategies in seconds.]{
\small{
\begin{tabular}{ | c | c | c | c |} 
\hline
\SetRowColor{MyGray} non-incr & manual-incr & efficient-incr & efficient-incr+storage \\
\hline 40957 & 10975 & 8007 & 6096\\
\hline
\end{tabular}
%\caption{Total runtime for SigmaClip in seconds }
\label{tab:SigmaClip-runtime} 
}
}
\caption{Runtime of the \texttt{SigmaClip} application with and without
incremental processing. Constant $k=3$ in all the algorithms.}
\vspace{-20pt}
\end{figure}

We first demonstrate the effectiveness of our approach to bringing
incremental processing to the iterative array model in the context of
the \texttt{SigmaClip} application.  Figure~\ref{fig:SigmaClip-incr}
shows the total runtime of the algorithm with different execution
strategies. As shown, the \texttt{non-incremental} ``sigma-clipping"
algorithm performs almost four times worse than any other
approach. The \texttt{manual-incr} approach is the
\texttt{incr-sigma-clipping} function from Section~\ref{sec:inc},
which is the manually-written incremental version of
the ``sigma-clipping" algorithm. This approach keeps track of all the
points that are still candidates to be removed at the next iteration
and discards the rest. By doing so, it touches the minimum number of
cells from the input dataset at each iteration. Although
\texttt{manual-incr} performs better than other approaches 
at later stages of the iterative computation, it incurs significant
overhead during the first few iterations due to the extra data points tracking 
(Lines~\ref{line:merge2} to~\ref{line:storage4} in \texttt{incr-sigma-clipping()} function).
%That is the reason why this
%approach is doing poorly overall, even worse than the non-incremental
%approach, at the first iteration. 
\texttt{manual-incr} also requires a \texttt{post-processing} phase at
the end of the iterative computation to return the final
result. \texttt{efficient-incr} and \texttt{efficient-incr+storage}
are the two strategies used by ArrayLoop (\texttt{ArrayLoop-incr-sigma-clipping} function from Section~\ref{sec:inc}). \texttt{efficient-incr} represents ArrayLoop's query rewrite for incremental state management that
also leverages our \texttt{merge} operator.
\texttt{efficient-incr+storage} further includes the storage manager
extensions. Figure~\ref{tab:SigmaClip-runtime} shows the total runtime
in each case. ArrayLoop's efficient versions of the algorithm are
competitive with the manually written variant. They even outperform
the manual version in this application.  All the incremental
approaches beat the non-incremental one by a factor of $4-6X$.
Interestingly, our approach to push some incremental computations to
the storage manager improves \texttt{efficient-incr} by an extra
$25\%$.

\subsection{Overlap Iterative Processing}
\begin{figure}[t]
\centering
\subfigure[\texttt{SourceDetect} application: T1, T5, and T10 refer to 
policies where ArrayLoop shuffles overlap data every iteration, every 5 iterations, and every 10 iterations, respectively. \texttt{converge}
is the strategy where ArrayLoop shuffles data only after local convergence occurs.]{
\includegraphics[width=0.80\linewidth]{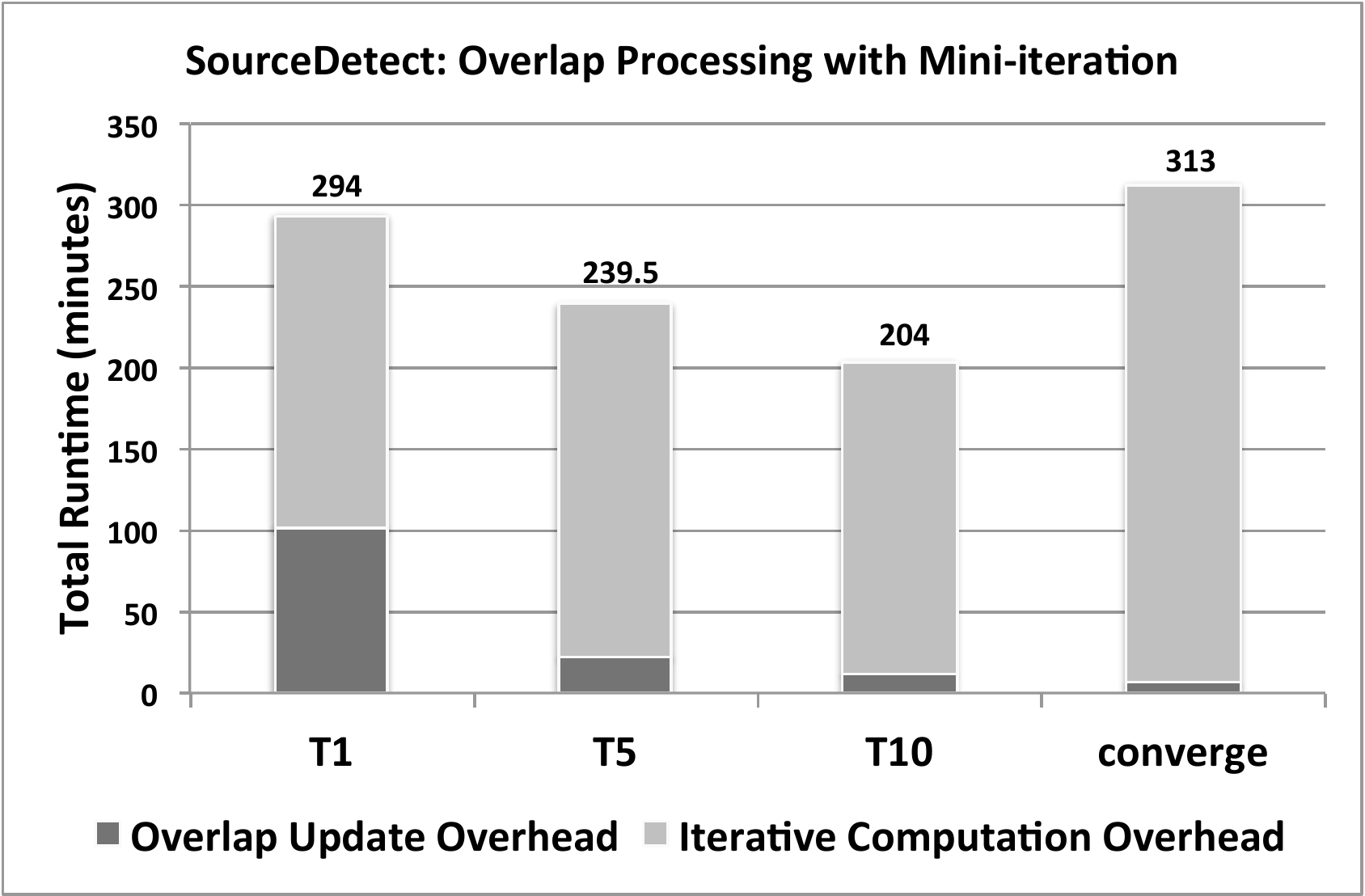}
\label{fig:detect-mini}
}
\subtable[Number of major and mini iterations. Major\# is the number of times that overlap data is reshuffled and Mini\# is the total number of iterations.]{
\scriptsize {
\begin{tabularx}{0.5\textwidth}{ |X | X | X | X | X |} 
\hline
\SetRowColor{MyGray} & T1 & T5 & T10 & converge \\
\hline Mini\# & 51 & 57 & 60 & 94\\
\hline Major\# & 51& 11 & 6 & 3 \\ 
\hline
\end{tabularx}
\label{tab:SourceDetect-mini}  
}
}
\caption{\texttt{SourceDetect} application: Iterative overlap processing with mini-iteration optimization.}
\vspace{-15pt}
\end{figure}

In Section~\ref{sec:overlap}, we describe overlap processing as a
technique to support \textit{parallel} array processing. In the case
of an iterative computation, the challenge is to keep the overlap data
up-to-date as the iteration progresses. The solution is to efficiently
shuffle overlap data at each iteration. An optimization applicable to
many applications is to perform \texttt{mini-iteration} processing,
where the shuffling happens only periodically.
Figure~\ref{fig:detect-mini} shows the effectiveness of this
optimization in the context of the \texttt{SourceDetect} application,
which requires overlap processing. \texttt{T1} refers to the policy
where ArrayLoop shuffles overlap data at each iteration, or no
\texttt{mini-Iteration} processing. As expected this approach incurs
considerable data shuffling overhead, although it converges faster in
the \texttt{SourceDetect} application
(Figure~\ref{tab:SourceDetect-mini}). At the other extreme, we
configure ArrayLoop to only shuffle overlap data after local convergence
occurs in all the chunks. Interestingly, this approach performs worse
than $T1$. Although this approach does a minimum number of data
shuffling, it suffers from the long tail of mini-iterations
(Figure~\ref{tab:SourceDetect-mini}: 94 mini-iterations). $T5$ and
$T10$ are two other approaches, where ArrayLoop shuffles data with some
constant interval. We find that $T10$, which shuffles data every ten
iterations, is a good choice in this application.  The
optimal interval is likely to be application-specific and tuning that value
automatically is beyond the scope of this paper. The other interesting
approach is to instruct ArrayLoop to initiate overlap data shuffling when
the number (or magnitude) of changes between mini-iterations is below
some threshold. We simply pick a constant number to determine the overlap data shuffling interval in the context of the \texttt{SourceDetect} application. More sophisticated approaches are left for future study.

\subsection{Multi-Resolution Optimization}
\begin{figure}[t]
\centering
  \subfigure[\texttt{SourceDetect} application: T10 refers to the
  strategy where ArrayLoop shuffles overlap data every 10 iterations. A0,
  A1, A2, A3, and A4 are five versions of the same array with
  different resolutions, where A0 is the same resolution as the
  original array and A4 is the most pixelated version. The grid-size
  is (2$\times$2).]{
\includegraphics[width=0.75\linewidth]{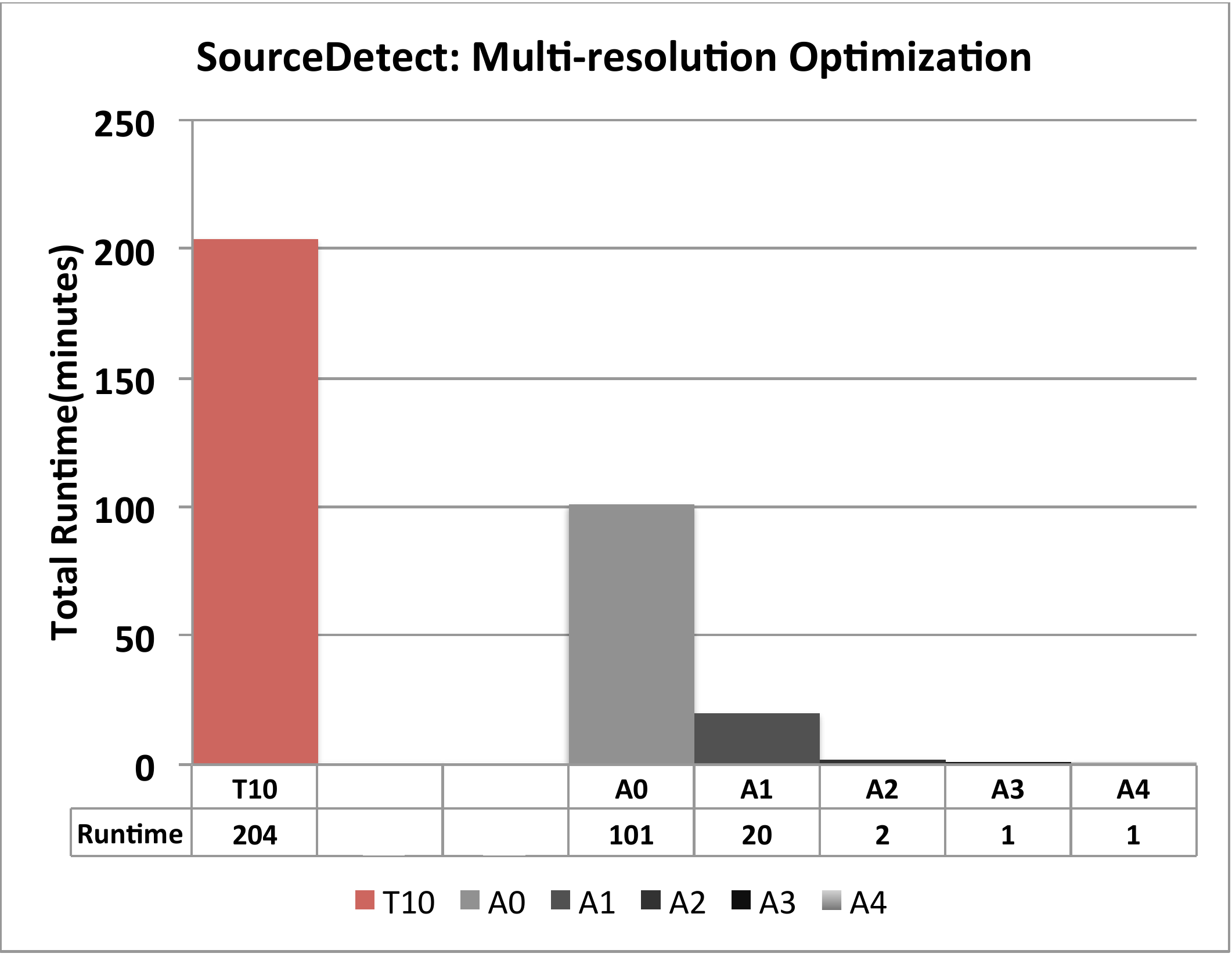}
\label{fig:multi-res}
}
\subtable[Iteration\# that converges at each resolution.]{
\scriptsize {
\begin{tabularx}{0.45\textwidth}{ |X |X | X | X | X | X | X |} 
\hline
\SetRowColor{MyGray} & T10 & A0 & A1 & A2 & A3 & A4 \\
\hline iter\# & 60 & 35 & 12 & 12 & 6 & 10\\
\hline
\end{tabularx}
\label{tab:multi-res}  
}
}
\caption{\texttt{SourceDetect} application: Multi-resolution Optimization.}
\vspace{-15pt}
\end{figure}

The \texttt{multi-resolution} optimization is a form of prioritized
processing. By first processing a low-resolution approximation of the
data, we focus on identifying the overall shape of the structures.
Further processing of higher-resolution (larger) arrays then extracts
the more detailed outlines of these
structures. Figure~\ref{fig:multi-res} shows the benefits of this
approach in the context of the \texttt{SourceDetect} application. We
generate four lower-resolution versions of the source array A0 by
sequentially calling the \texttt{grid()} operator with a grid-size
of (2$\times$2). We operate on these multi-resolution versions exactly
as described in Equation~\ref{eq:multi-res}. The performance
results are compared to those of \texttt{T10} from
Figure~\ref{fig:detect-mini} as we pick the same overlap-processing
policy to operate on each multi-resolution array. Interestingly, the
\texttt{multi-resolution} optimization cuts runtimes nearly in
half. Note that most of the saving comes from the fact that the
algorithm converges much faster in $A0$ compared to its counterpart
$T10$ (Figure~\ref{tab:multi-res}) thanks to the previous runs over
arrays $A1$ through $A4$, where most of the cell-points are already
labeled with their final cluster values. 

In Section~\ref{sec:multi-res}, we described a potential optimization
in case of input data changes in the original array. As an initial
evaluation of the potential of this approach, we modify the input data
by dropping one image from the large, 3D array. This change is
consistent with the LSST use-case, where a new set of images will be
appended to the array every night. We observe that the new co-added
image only differs in a small number of points from the original
one. Additionally, these changes do not affect the pixelated array
A1. This gives us the opportunity to re-compute the
\texttt{SourceDetect} application not from the beginning, but from the
pixelated version A1. Although the performance gain is not major in
this scenario, it demonstrates the opportunity for further novel
optimizations that we leave for future work.

\section{Related Work}
\label{sec:related}

Several systems have been developed that support iterative big data 
analytics~\cite{bu:10,ewen:12,low:12,shaw:12,zhang:11}.
Some have explicit iterations, others require an external driver to
support iterations, but none of them provides native support for iterative computation in the context of 
parallel array processing. 

Twister~\cite{ekanayake:10}, Daytona~\cite{daytona}, and HaLoop~\cite{bu:10} extend MapReduce 
to add a looping construct and preserve state across iterations. 
HaLoop takes advantage of the task scheduler to increase local access
to the static data. However, our system takes advantage of iterative \textit{array} processing 
to increase local access to the dynamic data as well by applying overlap iterative processing. 

%\emad{Spinning Fast Iterative Data Flows to be added} 

PrIter~\cite{zhang:11} is a distributed framework for fast iterative 
computation. The key idea of PrIter is to prioritize iterations that 
ensure fast convergence. In particular, PrIter gives each data point a priority value to indicate 
the importance of the update and it enables selecting a subset of data 
rather than all the data to perform updates in each iteration. ArrayLoop also supports a form of prioritized processing
through multi-resolution optimization. ArrayLoop initially finds course-grained outlines of the structures 
on the more pixelated versions of the array, and then it refines the details on fine-grained versions.

REX~\cite{rex} is a parallel shared-nothing query processing platform 
implemented in Java with a focus on supporting incremental iterative computations 
in which changes, in the form of deltas, are propagated from iteration 
to iteration. Similar to REX, ArrayLoop supports incremental iterative processing. However REX lacks other optimization techniques 
that we provide.

A handful of systems exist that support iterative computation with focus on graph algorithms. 
Pregel~\cite{malewicz:10} is a bulk synchronous message passing abstraction 
where vertices hold states and communicate with neighboring vertices.
Unlike Pregel, ArrayLoop relieves the synchronization barrier overhead by including mini-iteration 
steps in the iterative query plan. Unlike ArrayLoop, Pregel does not prioritize iterative computation. 

GraphLab~\cite{low:12} develops a programming model for iterative
machine learning computations. The GraphLab abstraction consists of
three main parts: the data graph, the dynamic asynchronous computation
as update functions, and the globally sync operation. Similar to our overlap
iterative processing technique, GraphLab has a notion of
\textit{ghost} nodes. However, the granularity of computation is per
node, while ArrayLoop supports overlap iterative processing per
chunk. Our system also supports prioritization through the novel
multi-resolution iterative processing.

Prior work also studies array processing on in-situ
data~\cite{blanas14}.  While this work addresses
the limitation that array data must first be loaded into an array
engine before it can be analyzed, it does not provide any special
support for iterative computation. SciDB and our extensions are
designed for scenarios where loading times amortize over a
sufficiently large amount of data analysis.

%Hyracks~\cite{borkar:11} is a parallel dataflow platform to run data-intensive
%jobs with a pipelined execution model over a DAG of abstract operators. 

\vspace{-5pt}
\section{Conclusion}

In this paper, we developed a model for iterative processing in a
parallel array engine.  We then presented three optimizations to
improve the performance of these types of computations: incremental
processing, mini-iteration overlap processing, and multi-resolution
processing. Experiments with a 1TB scientific dataset show that our
optimizations can cut runtimes by 4-6X for incremental processing,
31\% for overlap processing with mini-iterations, and almost 2X for
the multi-resolution optimization. Interestingly, the optimizations
are complementary and can be applied at the same time, cutting
runtimes to a small fraction of the performance without our approach.
\vspace{-5pt}
%\input{relwork}
%\vspace{-5pt}
\section*{Acknowledgments}

This work is supported in part by NSF grant IIS-1110370
and the Intel Science and Technology Center for Big Data. 
\end{sloppypar}

%******************************************************** References ****************************************************************
\balance
\vspace{2pt}
\small
\bibliographystyle{plain}
\bibliography{header,temporal,array-organization,parallel,distributed,scientific,self,recursive,graphs}

\begin{thebibliography}{10}

\bibitem{daytona}
R.~Barga et~al.
\newblock Daytona: Iterative {MapReduce} on {W}indows {A}zure.
\newblock
  \url{http://research.microsoft.com/en-us/projects/daytona/default.aspx}.

\bibitem{blanas14}
{Blanas et. al.}
\newblock Parallel data analysis directly on scientific file formats.
\newblock In {\em SIGMOD}, pages 385--396, 2014.

\bibitem{bu:10}
Y.~Bu et~al.
\newblock {HaLoop}: Efficient iterative data processing on large clusters.
\newblock {\em PVLDB}, 3(1), 2010.

\bibitem{curde:10}
{Cudre-Mauroux et. al.}
\newblock {SS-DB: A Standard Science DBMS Benchmark}.
\newblock
  \url{http://www-conf.slac.stanford.edu/xldb10/docs/ssdb_benchmark.pdf}, 2010.

\bibitem{ekanayake:10}
J.~Ekanayake et~al.
\newblock Twister: a runtime for iterative {MapReduce}.
\newblock In {\em HPDC}, pages 810--818, 2010.

\bibitem{soroush:11a}
E.Soroush, M.Balazinska, and D.Wang.
\newblock {ArrayStore}: A storage manager for complex parallel array
  processing.
\newblock In {\em SIGMOD}, pages 253--264, June 2011.

\bibitem{baumann:98}
Baumann et. al.
\newblock The multidimensional database system {RasDaMan}.
\newblock In {\em SIGMOD}, pages 575--577, 1998.

\bibitem{ewen:12}
S.~Ewen et~al.
\newblock Spinning fast iterative data flows.
\newblock In {\em VLDB}, pages 1268--1279, 2012.

\bibitem{furtado:99}
{Furtado et. al.}
\newblock Storage of multidimensional arrays based on arbitrary tiling.
\newblock In {\em Proc. of the 15th ICDE Conf.}, 1999.

\bibitem{gray:97}
{Gray, J. et. al.}
\newblock Data cube: A relational aggregation operator generalizing group-by,
  cross-tab, and sub-totals.
\newblock {\em dmkd}, pages 29--53, 1997.

\bibitem{hey:09}
T.~Hey, S.~Tansley, and K.~Tolle.
\newblock The fourth paradigm: Data-intensive scientific discovery, 2009.

\bibitem{astroMLText}
{\v Z}.~{Ivezi{\'c}}, A.J. {Connolly}, J.T. {Vanderplas}, and A.~{Gray}.
\newblock {\em Statistics, Data Mining and Machine Learning in Astronomy}.
\newblock Princeton University Press, 2014.

\bibitem{kwon:10a}
Y.~Kwon et~al.
\newblock Scalable clustering algorithm for {N-body} simulations in a
  shared-nothing cluster.
\newblock In {\em {SSDBM}}, 2010.

\bibitem{kwon:10b}
{Kwon, Y. et. al.}
\newblock Skew-resistant parallel processing of feature-extracting scientific
  user-defined functions.
\newblock In {\em Proc. of SOCC Symp.}, June 2010.

\bibitem{loebman:09}
{Loebman et. al.}
\newblock Analyzing massive astrophysical datasets: Can {Pig/Hadoop} or a
  relational {DBMS} help?
\newblock In {\em IASDS}, 2009.

\bibitem{low:12}
Y.~Low et~al.
\newblock Distributed {GraphLab}: a framework for machine learning and data
  mining in the cloud.
\newblock In {\em VLDB}, pages 716--727, 2012.

\bibitem{lsst}
{Large Synoptic Survey Telescope}.
\newblock \url{http://www.lsst.org/}.

\bibitem{mahout}
Apache mahout.
\newblock \url{http://mahout.apache.org/}.

\bibitem{malewicz:10}
G.~Malewicz et~al.
\newblock Pregel: a system for large-scale graph processing.
\newblock In {\em SIGMOD}, 2010.

\bibitem{mcsherry:13}
F.~McSherry, D.~G. Murray, R.~Isaacs, and M.~Isard.
\newblock Differential dataflow.
\newblock In {\em CIDR.}, 2013.

\bibitem{rex}
S.R. Mihaylov et~al.
\newblock {REX}: Recursive, delta-based data-centric computation.
\newblock In {\em VLDB}, 2012.

\bibitem{soroush:13b}
{Moyers et. al.}
\newblock A demonstration of iterative parallel array processing in support of
  telescope image analysis.
\newblock In {\em Proc. of the 39th VLDB Conf.}, 2013.

\bibitem{NietoSantisteban:06}
{Nieto-santisteban et. al.}
\newblock Cross-matching very large datasets.
\newblock In {\em NSTC NASA Conference}, 2006.

\bibitem{repository}
{UW-CAT}.
\newblock \url{http://myria.cs.washington.edu/repository/uw-cat.html}.

\bibitem{rogers:10}
J.~Rogers et~al.
\newblock Overview of {SciDB}: Large scale array storage, processing and
  analysis.
\newblock In {\em SIGMOD}, 2010.

\bibitem{scidbguide}
{SciDB Guide}.
\newblock \url{http://scidb.org/HTMLmanual/13.3/scidb_ug/ }.

\bibitem{scidb-py}
Scidb-py.
\newblock \url{http://jakevdp.github.io/SciDB-py/tutorial.html}.

\bibitem{seamons:94}
{Seamons et. al.}
\newblock Physical schemas for large multidimensional arrays in scientific
  computing applications.
\newblock In {\em Proc of 7th SSDBM}, pages 218--227, 1994.

\bibitem{shaw:12}
{Shaw et. al. }.
\newblock Optimizing large-scale semi-naive {D}atalog evaluation in {H}adoop.
\newblock In {\em In Datalog 2.0}, 2012.

\bibitem{skyserver}
{Sloan Digital Sky Survey} {III}: {SkyServer} {DR12}.
\newblock \url{http://skyserver.sdss.org/dr12/en/home.aspx}.

\bibitem{sdss}
{Sloan Digital Sky Survey}.
\newblock \url{http://cas.sdss.org}.

\bibitem{soroush:13a}
{Soroush et. al }.
\newblock Time travel in a scientific array database.
\newblock In {\em Proc. of the 29th ICDE Conf.}, March 2013.

\bibitem{taft:14}
{Taft et. al.}
\newblock Genbase: A complex analytics genomics benchmark.
\newblock In {\em SIGMOD}, pages 177--188, 2014.

\bibitem{zaharia:10}
M.~Zaharia et~al.
\newblock Spark: cluster computing with working sets.
\newblock In {\em HotCloud'10}, 2010.

\bibitem{zhang:11}
Y.~Zhang et~al.
\newblock {PrIter}: a distributed framework for prioritized iterative
  computations.
\newblock In {\em VLDB}, 2011.

\bibitem{zhang:09}
{Zhang et. al.}
\newblock {RIOT}: {I/O}-efficient numerical computing without {SQL}.
\newblock In {\em Proc. of the Fourth CIDR Conf.}, 2009.

\end{thebibliography}
\balance

\end{document}